\documentclass[11pt]{article}
\usepackage{amsmath,epsf,cite}
\usepackage{amssymb,amsthm,tocvsec2}
\usepackage{graphicx,float}
\usepackage{subfigure}
\usepackage{setspace}
\usepackage{mdwlist}
\usepackage{paralist}
\usepackage{setspace}
\usepackage{enumitem}
\usepackage{indentfirst}
\usepackage{epstopdf}
\usepackage{empheq}
\usepackage{grffile}
\usepackage{color}
\usepackage{CJK}
\usepackage{verbatim}
\usepackage[margin=1in]{geometry}

\date{\today}

\numberwithin{figure}{section}
\numberwithin{table}{section}
\numberwithin{footnote}{section}

\theoremstyle{plain}

\newtheorem*{thm*}{Theorem}

\theoremstyle{remark}

\numberwithin{equation}{section}

\def\bn{\mathbf{n}}
\def\bv{\mathbf{v}}
\def\bx{\mathbf{x}}

\def\bz{\mathbf{z}}

\def\bA{\mathbf{A}}

\def\bv{\mathbf{v}}
\def\bh{\mathbf{h}}

\def\bR{\mathbb{R}}

\newcommand{\parl}[2]{\ensuremath{\frac{\partial #1}{\partial #2}}}
\newcommand{\parls}[2]{\ensuremath{\frac{\partial^2 #1}{\partial #2^2}}}
\newcommand{\parlm}[3]{\ensuremath{\frac{\partial^2 #1}{\partial #2 \partial #3}}}
\newcommand{\bena}{\begin{eqnarray}\begin{array}{l}}
\newcommand{\eena}{\end{array}\end{eqnarray}}
\newcommand{\ben}{\begin{eqnarray}}
\newcommand{\een}{\end{eqnarray}}
\newcommand{\beq}{\begin{equation}}
\newcommand{\eeq}{\end{equation}}
\newcommand{\bea}{\begin{array}}
\newcommand{\eea}{\end{array}}
\newcommand{\bef}{\begin{figure}}
\newcommand{\eef}{\end{figure}}

\def\begd#1\eegd{\vskip 0.3\baselineskip $\begin{aligned}#1\end{aligned}$\vskip 0.3\baselineskip}

\begin{document}

\title{     Boundary-to-Solution Mapping for Groundwater Flows in a Toth Basin }

{\tiny
\author{ Jingwei Sun${}^{\dagger}$,   Jun Li${}^{\dagger\dagger}$, Yonghong Hao${}^{\diamond}$, Cuiting Qi${}^{\ddagger\ddagger}$, Chunmei Ma${}^{\dagger}$, Huazhi Sun${}^{\dagger}$,\\
Negash Begashaw${}^{\diamond\diamond}$,
Gurcan Comet${}^{\diamond\diamond}$,  Yi Sun${}^{\ddagger}$, and  Qi Wang${}^{\ddagger}$  \\
$\dagger$
School of Computer Science, Tianjin Normal University, Tianjin, China,100084;\\
$\dagger\dagger$
School of Mathematical Science, Tianjin Normal University, Tianjin, China, 300387;\\
$\diamond$
Center for Ground Water Research, Tianjin Normal University, Tianjin, China,100084;\\
$\ddagger\ddagger$
School of Geographical and Environmental Science, Tianjin Normal University\\
 Tianjin, China,100084;\\
$\diamond\diamond$
Mathematics, Computer Science, Physics and Engineering Department, Benedict College,\\
 Columbia, SC 20204, USA;\\
$\ddagger^{*}$
Department of Mathematics, University of South Carolina, \\
Columbia, SC 29208, USA; Email: qwang@math.sc.edu.
 }}
\date{\today}
\maketitle

\begin{abstract}

In this paper, the authors propose a new approach to solving the groundwater flow equation in the Toth basin of arbitrary top and bottom topographies using deep learning. Instead of using traditional numerical solvers, they use a DeepONet to produce the boundary-to-solution mapping. This mapping takes the geometry of the physical domain along with the boundary conditions as inputs to output the steady state solution of the groundwater flow equation.
To implement the DeepONet, the authors approximate the top and bottom boundaries using truncated Fourier series or piecewise linear representations. They present two different implementations of the DeepONet: one where the Toth basin is embedded in a rectangular computational domain, and another where the Toth basin with arbitrary top and bottom boundaries is mapped into a rectangular computational domain via a nonlinear transformation. They implement the DeepONet with respect to the Dirichlet and Robin boundary condition at the top and the Neumann boundary condition at the impervious bottom boundary, respectively.
Using this deep-learning enabled tool, the authors investigate the impact of surface topography on the flow pattern by both the top surface and the bottom impervious boundary with arbitrary geometries. They discover that the average slope of the top surface promotes long-distance transport, while the local curvature controls localized circulations. Additionally, they find that the slope of the bottom impervious boundary can seriously impact the long-distance transport of groundwater flows.
Overall, this paper presents a new and innovative approach to solving the groundwater flow equation using deep learning, which allows for the investigation of the impact of surface topography on groundwater flow patterns.

\end{abstract}


\section{Introduction}
The Toth groundwater flow analysis was a seminal theoretical attempt to relate surface topography of the water table and the associated hydrological boundary conditions with the steady state groundwater flow field driven by gravity in a small drainage basin, known as the Toth basin \cite{tothTheoryGroundwaterMotion1962,Toth1963}. It involved solving an elliptic boundary value problem for a given surface topography of the water table not far from a horizontally flat surface with the associated Dirichlet boundary condition on a rectangular domain approximately.  For a general non-rectangular drainage basin with a surface topography far from a flat surface, the elliptic boundary value problem would have to be  solved numerically. The Toth water table analysis demonstrated the impact of surface topography and the associated water potential at the boundary on the ground water flow in the basin domain  approximately.  Traditionally, a numerical solver (such as MODFLOW, COMSOL, etc.) for the solution of groundwater flow equations only produces one solution for any given boundary conditions. When one studies another Toth basin with a different surface topography, the solution have to be recalculated completely. Given the flow equation and the boundary condition, the mapping from the boundary condition to the solution is essentially provided by the numerical solver. One thus wonders if  the numerical solver can be replaced by a concrete function or "mapping" that is fully  capable of  producing the solution from any prescribed surface topography without  being recalculated.

In the past,  the Toth theory was refined through adjusting the coefficient of permeability or the viscosity of the fluid in porus media, extended to examine the influence of temperature \cite{2014ARan}, and used to investigate the influence of depth and systemic heterogeneity in porus media \cite{2010Groundwater,Jiang&W2011}. Several studies focused on generalizing the Toth theory to other settings, like the more realistic three-dimensional space \cite{2017IdentifyingWang}, unsteady situations \cite{2015AnalyticalNiu}, etc. However, none of the studies paid attention to the impact of the top surface topography of the water table on the solution in the Toth basin holistically when it is of an arbitrary shape.

In this paper, we extend the Toth water table study to a domain with an arbitrary piecewise smooth top and bottom boundaries and two physically relevant boundary conditions, and propose a novel approach to establish a mapping from surface topographies (top alone or top+ bottom) to the solution of the groundwater flow equation  in the Toth basin directly using a deep learning approach. To some extent, this is an analogue of a solution formula for an initial-boundary value problem for partial differential equations (PDEs) in the context of deep learning,  where the solution of the initial-boundary value problem  is expressed as a neural network function of the domain, the boundary conditions and the nonhomogeneous forcing term. This approach produces a solution mapping that maps the prescribed initial and boundary conditions as well as the forcing term to the solution directly. It can be readily applied to any geophysical basins that share the same hydrological property such as the mobility/conductivity coefficient in the flow equation.
As a demonstration of the approach, we present the mapping while neglecting the forcing effect due to the source or sink and assuming the porus media is spatially homogeneous. We remark that the method applies to any inhomogeneous porus media and groundwater flow equations with a source or sink term. The advantage of this approach is that once the solution mapping is obtained in one Toth basin, it can be readily applied to all other Toth basins where the hydrological property of the porus media is the same, but the boundary and boundary conditions can be different.

The recent advancement in deep learning with neural networks makes the development of such a  desired mapping plausible \cite{karniadakisPhysicsinformedMachineLearning2021a,Lu2019DeepONetLN, caoChooseTransformerFourier,guibasADAPTIVEFOURIERNEURAL2022,kissasLEARNINGOPERATORSCOUPLED2022,
liFourierNeuralOperator2021,pangFPINNsFractionalPhysicsInformed2019}. Given that a neural network is a mapping composed of compound functions with specific layered structures, the mapping can be established should we propose the proper architecture of the deep neural network in principle in the context of physics-informed machine learning (PIML) \cite{leshnoMultilayerFeedforwardNetworks1993,1995Universal}. We note  that the  steady state groundwater flow equation  in porus media is an elliptic (or Poisson) equation. Given the boundary and physically consistent boundary conditions, a solution can be represented by an integral containing the Green's function \cite{habermanAppliedPartialDifferential2013}. The integral with the Green's function yields the mapping from the boundary, boundary conditions and the source term  to the solution theoretically. Motivated by this connection between the domain, boundary conditions and the source term of the equation, we represent the mapping  using a new form of neural network, known as the DeepONet. The DeepONet has been shown to have the capacity to establish the mapping between the model parameters,  its boundaries (including boundary conditions)  to the solution in the domain \cite{Lu2021LearningNO}. It is therefore an appropriate and powerful tool for us to build the desired boundary-to-solution  mapping.

Specifically, we will answer the following questions in this study using machine learning with DeepONet.
\begin{itemize}
\item 	What is the influence of the surface topography and the geometry of the Toth basin to the steady flow field through the water potential in Toth basin $\Omega$?
\item 	What is the specific effect of both the top and bottom boundary conditions to the solution of the groundwater flow equation in the Toth basin through the boundary-to-solution mapping? We will focus on two types of top boundary conditions: (i) the Dirichlet boundary condition in which the water potential is prescribed at the top boundary related to the altitude of the location: $h=g\phi(\bx)$, where $h$ is the water potential, $g$ is gravity, and $y=\phi$ defines the top boundary; and (ii) the Robin boundary condition: $ \frac{\partial h}{\partial \bn}(\bx)+\gamma h(\bx)=\gamma g\phi(x)$, where $\gamma$   is a rate parameter whose reciprocal represents the penetration length. The latter simply states a balance law between the cross boundary flux and the difference between the water potential and a saturated water potential at the top surface. As $\gamma \to \infty$, i.e. the penetration length shrinks to zero, the Dirichlet boundary condition is recovered. Thus, the Robin boundary condition is an approximation to the Dirichlet boundary condition at large $\gamma>>1$.
    \item What's the surface topography and basin geometry to the steady state Darcy velocity field in the basin? We note that the topography here refers to the water table topography or water head profile not the topography of the ground surface.
\end{itemize}
We will address these issues holistically by solving the PDE boundary value problem with respect to two distinct boundary conditions using DeepONet  \cite{Pang2019fPINNsFP, Lu2019DeepONetLN}. The presentation is given for steady states without a source term. However, we emphasize again that the method extends readily to flow phenomena with a source and transient situations with a minimum modification to the DeepONet architecture. We note that for a completely new Toth basin, an analogous boundary-to-solution mapping to describe ground water flows in the porous medium can be obtained through the transfer learning, which could accelerate the training process and be efficiently done.

Numerically, we present three implementations of the DeepONet in which the boundaries of arbitrary shapes are represented using   a piecewise linear interpolant, a truncated Fourier series, or mapped  to a flat surface via a nonlinear transformation. All three implementations yield comparable numerical results. Without loss of generality, we will detail the latter two implementations in this paper.



\section{Mathematical formulation }

 We first present the model derivation and give a brief discussion on consistency of boundary conditions with the governing equation. Then, we discuss how the solution of the boundary value problem of the steady state groundwater flow equation depends on prescribed boundary conditions and source to set up the stage to apply physical-informed-machine-learning (PIML) with neural networks to solve the boundary value problem.

\subsection{Model formulation}

We formulate the groundwater flow model in a general time-dependent setting. We consider flow  of ground water in a given domain $\Omega$ with piecewise smooth boundary $\partial \Omega$, in which some parts are impervious. We denote the water potential by $h(\bx, t)$ at location $\bx$ and time $t$. It is related to the hydrostatic pressure through
\ben
h(\bx, t)=gy+\int_{p_0}^p \frac{1}{\rho} dp,\label{def-h}
\een
where $g$ is the gravity acceleration, $y$ is the height of the water basin measured from the bottom impervious layer,  $\rho(p)$ is the density of water, a function of pressure $p$, $p_0$ is the atmospheric pressure at the top surface of the water table and $p(\bx,t)$ is the hydrostatic pressure at $\bx$. Since the water potential is a gauge variable, we choose the origin of the coordinate system at the lower impervious layer so that the water potential at the surface is determined by the altitude of the top surface relative to the impervious layer. We remark that the origin for $y$ is chosen as the lowest point along the bottom surface when it is not flat.  The flow equation of $h(\bx,t)$ is given by the following
	continuity equation:
\ben\label{eq1}
S\frac{\partial h}{\partial t}=\nabla \cdot \bv+Q,
\een
where  $S$  is the storage rate, $Q$ is the source term, $\bv$ is the effective velocity or the Darcy velocity. It follows from \eqref{def-h} that
\ben
\nabla h=\nabla (gy)+\frac{1}{\rho}\nabla p.
\een

The constitutive equation between water potential $h$ and Darcy velocity $\bv $ is given by the Darcy's law\cite{nieldConvectionPorousMedia2006}:
\ben\label{eq2}
\bv =K \cdot \nabla h,
\een
where  $K$ is the mobility or conductivity coefficient tensor. We note that \eqref{eq2} can be viewed as a force balance equation, where the inverse, $K^{-1}$, serves as the friction coefficient. It follows from \eqref{eq1} and \eqref{eq2} that
\ben
S\frac{\partial h}{\partial t}=\nabla \cdot (K\cdot \nabla h)+Q.
\label{ge_1}
\een
This is the governing equation for water potential $h$ from which the Darcy velocity is inferred.

\subsection{Dirichlet boundary-value problem}

\begin{figure}[t]\label{Fig1}
\centering

			\includegraphics[width=0.7\textwidth]{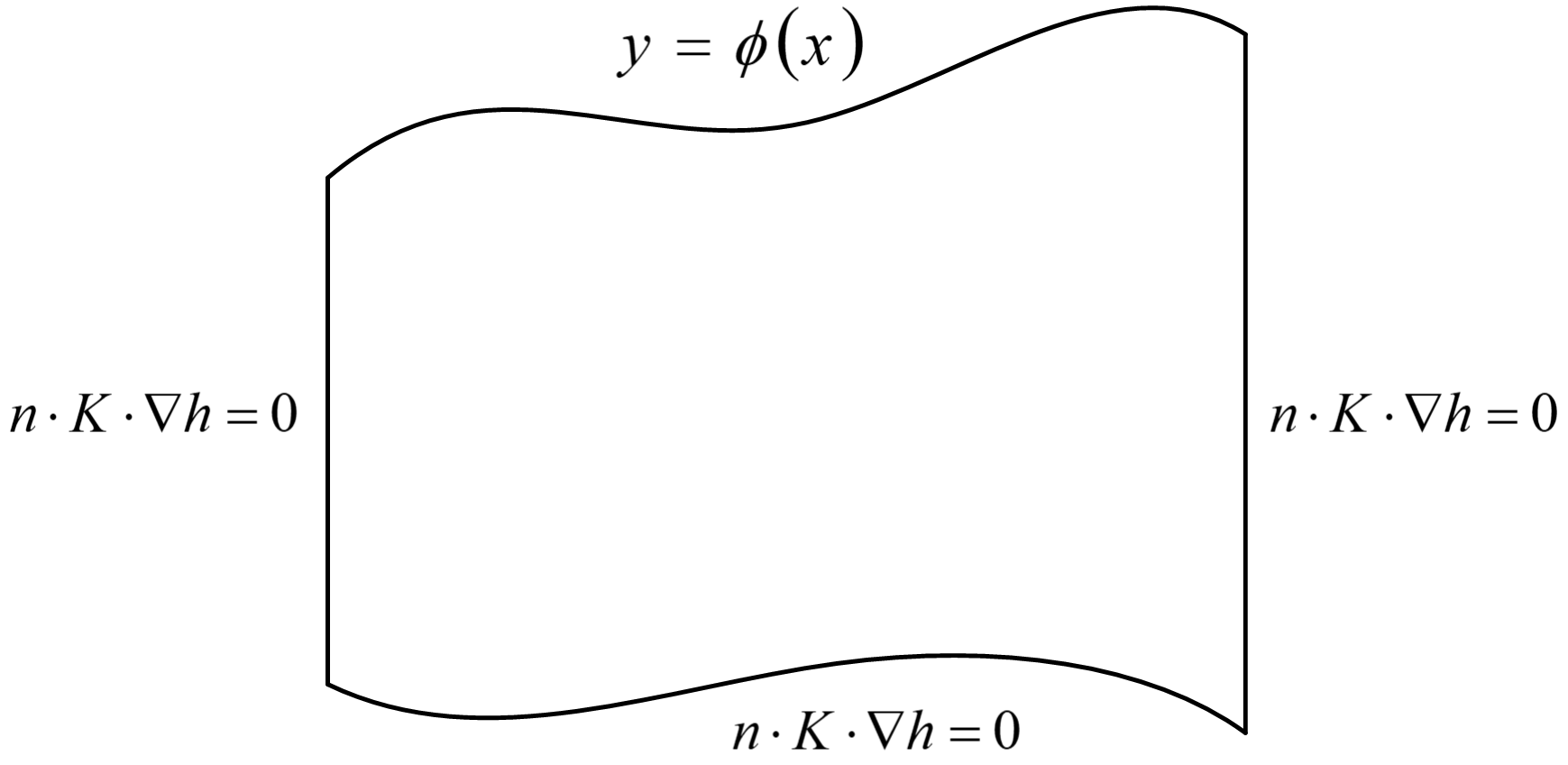}

\caption{Toth basin $\Omega$ and the prescribed boundary conditions over $\partial \Omega$ embedded in a rectangular domain $[0,1]\times [0,1]$. The top and bottom boundaries are given  by $y=\phi(\bx)$ and $ y=\phi_2(\bx)$, respectively.}
\end{figure}

In a water basin $\Omega$, this partial differential equation is accompanied by a set of boundary conditions over domain boundary $\Gamma=\partial \Omega$. We consider the following 2D domain   with boundary conditions given below (see Figure \ref{Fig1}),
\ben
\bn\cdot K\cdot \nabla h|_{\Gamma_{b,l,r}}=0, \quad h(x, \phi(x))|_{\Gamma_u}=g\phi(x), \label{BC}
\een
 where $\bn=\frac{1}{\sqrt{1+\phi_x^2}}(-\phi_x, 1)$ is the unit external normal to the boundary, $\Gamma_{b,l,r,t}$ are the boundaries at the bottom, left,  right and top side of domain $\Omega$, respectively,  and equation $y=\phi(x)$ defines the top boundary ($\Gamma_t$). The lateral boundaries are assumed vertical line segments in domain $\Omega$ while the top and bottom ones can be arbitrary. We name this domain the Toth basin for its origin in the Toth's seminal paper on the Toth water table. Notice that the lateral boundaries and the bottom one are assumed impervious in the Toth basin while the top one is not \cite{Toth1963}. When the bottom  boundary is flat and top boundary inclined with a small slope, Toth calculated his well-known Toth water table solution in \cite{Toth1970} using an approximate analytical method based on an asymptotic analysis on a rectangular domain.

Given any boundary conditions along $\partial \Omega$, we need to check their consistence with the governing equation in $\Omega$\cite{liSecondOrderLinear}. We integrate equation \eqref{ge_1} over $\Omega$ to obtain
\ben
\int_{\Omega} [S\frac{\partial h}{\partial t}-(\nabla \cdot (K\cdot \nabla h)+Q)]d\bx=\int_{\Omega} [S\frac{\partial h}{\partial t}-Q]d\bx-\int_{\partial \Omega} \bn \cdot (K\cdot \nabla h) ds =0.
\een
It imposes a consistent condition between the boundary conditions on $h$ and the solution in the interior. If boundary conditions are given in \eqref{BC}, the consistent condition reduces to
\ben
\int_{\Omega} [S\frac{\partial h}{\partial t}-Q]d\bx-\int_{ \Gamma_{t}} \bn \cdot (K\cdot \nabla h) ds =0.\label{constr0}
\een
In steady states and without the source term, in particular, the consistent condition further reduces to
\ben
\int_{ \Gamma_{t}} \bn \cdot (K \cdot \nabla h) ds =0.\label{constr1}
\een
The consistent condition is a crucial constraint for the equation to have a steady state solution.   Physically, this condition indicates that the net-flux across the top boundary in steady state must be zero. For the given top boundary $y=\phi(x)$, the unit external normal $\bn $ couples $\phi$ to
 solution $h(\bx)$ obtained in $\Omega$ through  \eqref{constr1}.

We summarize the mixed boundary value problem with the Dirichlet boundary condition on the top as follows
\ben
\left \{
\bea{l}
\nabla \cdot K\cdot \nabla h=0, \bx \in \Omega, \\
\bn\cdot K\cdot \nabla h|_{\Gamma_{b,l,r}}=0, \quad h(x, \phi(x))|_{\Gamma_t}=g\phi(x).
\eea \right. \label{IB}
\een
Assuming the boundary-value problem is well-posed, $h$ is a solution of \eqref{IB}, and $\hat{h}$ is another function of the same regularity as $h$, $\hat{h}$  satisfies the following estimate:
\ben
\|\hat{h}-h\|_{\Omega}\leq C_1\|\nabla \cdot K\cdot\nabla \hat{h}\|_{\Omega}+C_2 \|\bn\cdot K\cdot \nabla \hat{ h}\|_{\Gamma-\Gamma_t}+C_3 \|\hat{h}-g\phi\|_{\Gamma_t},
\een
where the norms are some proper norms defined in their respective spaces and $C_i, i=1,2,3$ are positive constants \cite{Evans2010PartialDE}.
Then,
\ben
h=arg \min_{\hat{h}} \|\hat{h}-h\|_{\Omega}=arg\min_{\hat{h}}[C_1\|\nabla \cdot K\cdot\nabla \hat{h}\|_{\Omega}+C_2 \|\bn\cdot K\cdot \nabla \hat{ h}\|_{\Gamma-\Gamma_t}+C_3 \|\hat{h}-g\phi\|_{\Gamma_t}].
\een
Thus, we  use the righthand side to define the loss function in this case.
\ben
Loss=C_1\|\nabla \cdot K\cdot\nabla \hat{h}\|_{\Omega}+C_2 \|\bn\cdot K\cdot \nabla \hat{ h}\|_{\Gamma-\Gamma_t}+C_3 \|\hat{h}-g\phi|_{\Gamma_t}.\label{Loss-D}
\een
In this case, finding the solution of \eqref{IB} is turned into a minimization problem of the residues in \eqref{Loss-D}. This is the foundation of PIML formulation \cite{karniadakisPhysicsinformedMachineLearning2021a}. The crucially important
part in this formulation is the choice of the norms in the loss function so that it is consistent with the well-posedness proof of the initial boundary value problem \cite{RAISSI2019686}.
In practice, we augment the loss function defined in \eqref{Loss-D} by a penalization of the consistent condition given in \eqref{constr1} as follows
\ben
Loss=C_1\|\nabla \cdot K\cdot\nabla \hat{h}\|_{\Omega}+C_2 \|\bn\cdot K\cdot \nabla \hat{ h}\|_{\Gamma-\Gamma_t}+C_3 \|\hat{h}-g\phi|_{\Gamma_t}+L_1(\int_{ \Gamma_{t}} \bn \cdot (K \cdot \nabla h) ds)^2,\label{Loss-DD}
\een
where $L_1>0$ is a model parameter set by the user. In this paper, we set $L_1=1$.

 \subsection{Robin boundary-value problem}

A more physical boundary condition in steady states of the groundwater flow equation  at the top boundary perhaps should be
\ben
\bn \cdot K\cdot \nabla h=-\gamma (h-g\phi(x)), \label{BC3}
\een
where $\gamma$ is the rate parameter. It indicates that the flux through the top boundary is proportional to the difference of the water potential and the saturated steady state water potential. If $\gamma=0$, \eqref{BC3} reduces to the impervious Neumann boundary condition; whereas it reduces to the Dirichlet one if $\gamma \to \infty$.

If we assume that the Robin boundary-value problem is well-posed, $h$ is a solution, and $\hat{h}$ a function in $H^1(\Omega)$, it follows from the well-posedness that
\ben
\|h-\hat{h}\|\leq C_1 \|\nabla \cdot K\cdot \nabla \hat{h}\|_{\Omega}+C_2\|\bn \cdot K\cdot \nabla \hat{h}\|_{\Gamma-\Gamma_t}+C_3 \|\bn \cdot K\cdot \nabla h+\gamma (\hat{h}-g\phi(x))\|_{\Gamma_t},
\een
where $C_i, i=1,2,3$ are positive constants. The loss function can then be devised as follows
\ben
Loss=C_1\|\nabla \cdot K \cdot \nabla \hat{h}\|_{\Omega}+C_2 \|\bn\cdot K\cdot \nabla \hat{ h}\|_{\Gamma-\Gamma_t}+C_3 \|\bn \cdot K\cdot \nabla h+\gamma (\hat{h}-g\phi(x))\|_{\Gamma_t}.\label{Loss-R}
\een
This loss function penalizes all the residues in the equation and the boundary conditions.

 The steady state governing equation  without the source together with boundary condition \eqref{BC3} yields the following consistency condition:
\ben
\int_{\Gamma_t} \gamma (h-\phi) ds=0. \label{consistent-2}
\een
\eqref{constr1} and \eqref{consistent-2} are two constraints for the solution to satisfy  the Dirichlet and the Robin boundary condition, respectively, which must be ensured in any solution solvers.
In practice, the loss function used in machine-learning in this study is  given by
\ben
Loss=C_1\|\nabla \cdot K \cdot \nabla \hat{h}\|_{\Omega}+C_2 \|\bn\cdot K\cdot \nabla \hat{ h}\|_{\Gamma-\Gamma_t}+C_3 \|\bn \cdot K\cdot \nabla h+\gamma (\hat{h}-g\phi(x))\|_{\Gamma_t}+(\int_{\Gamma_t} \gamma (h-\phi) ds)^2.\label{Loss-RR}
\een
\subsection{Nondimensionalizaton}

In order to solve the equations together with the boundary conditions numerically, we need to nondimensionalize them. We introduce length scale in x: $L_x$,  in y: $L_y$, and time scale: $T$, respectively.
The dimensionless variables are defined as follows
\ben
\tilde x=\frac{x}{L_x}, \tilde y=\frac{y}{L_y}, \tilde t=\frac{t}{T}, \tilde h=\frac{h}{h_0}, \tilde {\phi}=\frac{\phi}{L_y},
\een
where $h_0$ is a characteristic water potential.
The top Dirichlet boundary condition is given  by
\ben
\tilde h=\frac{g L_y}{h_0} \tilde \phi.
\een
We denote the characteristic storage rate by $S_0$. The dimensionless model parameters are given by
\ben
\tilde S=\frac{S}{S_0}, \tilde K=\frac{T}{L_y^2 S_0} \bA \cdot K \cdot \bA, \tilde Q=\frac{TQ}{S_0 h_0},
\een
where
\ben
\bA=
\left (
\bea{lr}
\epsilon &0\\
0 &1
\eea\right),
\een
$\epsilon=\frac{L_y}{L_x}$ is the aspect ratio of the basin.
The flow equation in the dimensionless form is given by
\ben
\tilde S \frac{\partial \tilde h}{\partial \tilde t}=\nabla \cdot \tilde K \cdot \nabla \tilde h+\tilde Q.
\een

We choose
\ben
h_0=gL_y,
\een
and drop the $\tilde{}$ from the dimensionless equations to obtain the dimensionless equation and boundary conditions as follows:
\ben
\left \{
\bea{l}
 S \frac{\partial  h}{\partial  t}=\nabla \cdot  K\cdot  \nabla  h+ Q, \bx \in \Omega,
 \\
  h= \phi(x), \bx \in \Gamma_{t}, \quad \bn \cdot K \cdot \nabla h=0, \bx \in \partial \Omega-\Gamma_{t}.
  \eea\right.
\een
The consistent condition \eqref{constr1} retains.

Analogously, we obtain the dimensionless Robin boundary condition at the top boundary as follows
\ben
\bn \cdot K \nabla  h=-\tilde \gamma({h}-\phi),
\een
where $\tilde \gamma=\gamma L_x.$ We drop the tilde over $\gamma$ for brevity in the following.
In this paper, we consider $K=Diag(K_{11}, K_{22})$ as a diagonal mobility matrix in the dimensionless equation, $x\in [0,1]$,  and  $0<\phi_2(\bx)< \phi(\bx)\leq 1$  as bottom and top boundaries, where $y=\phi_2$ represents the bottom boundary.

Next, we present three  implementations of the DeepONet  for the mapping from specified boundaries and the boundary conditions \eqref{BC}   to the steady state  solution of the groundwater flow equation  in $\Omega$ \cite{Lu2021LearningNO}, from which the Darcy's velocity can be recovered.

\section{DeepONet for the boundary-to-solution mapping}

 For the boundary value problem in the Toth basin, we would like to establish a mapping from the boundary and the associated boundary condition to the solution of the steady state groundwater flow equation in the domain. We adopt the physics-informed machine learning approach  and use the DeepONet as the neural network to represent the mapping  \cite{Lu2021LearningNO}. For the Toth basin, we consider a domain with flat lateral, arbitrary top and bottom boundaries. Firstly, we  consider a domain of aspect ratio $\epsilon$ with flat and impervious bottom and lateral boundaries   and prescribed water potential at the top boundary, $y=\phi(x)$.   We construct the top surface   to the solution mapping in the Toth basin with $(\phi, \epsilon)$ as the input. Owing to the fact that the dimensionless boundary condition coincides with the boundary representation, we only need to  learn a mapping from top boundary $y=\phi(x)$ with aspect ration $\epsilon$ to the solution in $\Omega$. We present three different approaches to accomplishing this goal using two distinct representations of top boundary $y=\phi(x)$, respectively. Secondly, we  discuss an extension of  the approach to the domain where the bottom boundary and the top boundary are both arbitrary.

\subsection{Piecewise polynomial interpolation of the top boundary}

We represent top boundary $y=\phi(x)$  using $n$ discrete points $\bx_i=(x_i, \phi(x_i)), i=1,\cdots, n$ uniformly distributed in the x-coordinate, where $x_i=(i-1)\Delta x, \Delta x=\frac{1}{n-1}$. We acknowledge a new development in treating the boundary condition in a weak formulation of partial differential equations by introducing new variable to satisfy the homogeneous boundary conditions in PIML \cite{Sukumar&S2021}. However, our proposed approach suffices for the current problem.  We denote the approximate solution of this mixed  boundary value problem in the interior of $\Omega$ by a DeepONet $G(\bh_t, \epsilon, \bx)$ as follows
\ben
G(\bh_t,\epsilon, \bx)=\sum_{k=1}^p\sum_{i=1}^q c_{i}^k\sigma(\sum_{j=1}^n \xi_{ij}^k \phi(\bx_j)+\xi_{i0}^k \epsilon+\theta_i^k)\sigma(W_k\cdot \bx+\zeta_k)+b_0,
\een
where $c_i^k$, $\xi_{ij}^k, W_k$ are weights and $\theta_i^k, \zeta_k, b_0$ are biases of the neural network, $\bh_t=(\phi(x_1), \cdots, \phi(x_n)) \in \bR^n$ denotes the uniformly distributed, y-coordinates of the interpolating points at the top boundary, and $n,p,q$ are positive integers.
The DeepONet represents the mapping from $\bh_{t}, \epsilon$ to the solution.

To apply the PIML method to learn the neural network, we choose $n_{l}, n_{r}, n_{b}, n_t, n_{i}$ points at the left, right, bottom, and top boundary, and the interior randomly.
For convenience, we use odd number for $n_{t}=n$  at the top boundary.
The loss function of the machine learning model is given by \eqref{Loss-DD} with the $L_2$ norms in the interior and on the boundary. We evaluate the integral norms using the Monte Carlo sampling. For  the randomly chosen points along the boundary and in the interior, $\{\bx_j^i, i=l,r,b, \bx_j\}$, and a well-defined uniform division of $[0,1]$, $\{x_j, j=1,\cdots, n_t \}, \bx_j^t=(x_j, \phi( x_j))$, the specific expression of the loss function is given by
\ben
\bea{l}
L(\theta, \bh_t, \epsilon)=\frac{1}{n}\sum_{i=1}^{n}(\phi(\bx_i)-G(\bh_t,\epsilon,\bx_i^t))^2
+\frac{1}{n_{i}}\sum_{j=1}^{n_{i}}(\nabla\cdot K\cdot\nabla G(\bh_t,\epsilon,\bx_j))^2
+\\\\
\frac{1}{n_{l}}\sum_{i=1}^{n_{l}}(\frac{\partial}{\partial x} G(\bh_t,\epsilon, \bx_i^l))^2
+
 \frac{1}{n_{r}}\sum_{i=1}^{n_{r}}(\frac{\partial }{\partial x}  G(\bh_t,\epsilon, \bx_i^r))^2
 +
\frac{1}{n_{b}}\sum_{i=1}^{n_{b}}(\frac{\partial }{\partial y} G(\bh_t, \epsilon, \bx_i^b))^2\\
\\
+ [\frac{\Delta x}{3}\big(F(\bh_t, \epsilon, \bx_1)
+4\sum_{i~even \in\{ 2,\cdots,n-1\}}F(\bh_t, \epsilon,\bx_i^t) +2\sum_{i~odd \in\{ 3,\cdots,n-2\}} F(\bh_t, \epsilon,\bx_i^t) +\\\\
F(\bh_t, \epsilon,\bx_n^t) \big)]^2,
\eea
\een
where
the boundary mass flux is given by
\ben
\bea{l}
F(\bh_t, \epsilon,\bx)=\bn^{(t)}\cdot K\cdot ( \frac{\partial}{\partial x}G(\bh_t, \epsilon,\bx),\frac{\partial }{\partial y}G(\bh_t,\epsilon, \bx))^T=-K_{11}\phi_x(\bx) G_x(\bh_t, \epsilon, \bx)+\\\\
K_{22}G_y(\bh_t, \epsilon, \bx).
\eea
\een
 The Simpson's quadrature formula is employed \cite{butcherNumericalMethodsOrdinary2016} to ensure the integral is accurate up to the fourth order in $\Delta x$. This is the PIML formulation of the problem where the residues in the equation and boundary conditions are penalized in the loss function in the $L_2$ norm. This loss function is defined for each given top boundary parameterized by $\bh_t$ and a set of randomly selected points from other parts of the domain. We note that this loss also includes a penalization term for constraint \eqref{constr1} to enforce consistency.

In the practical implementation, we  modify the loss function by re-balancing the weights.  The  loss function used in machine-learning is then modified into
 \ben
\bea{l}
L(\theta,\bh_t, \epsilon)=\frac{\lambda_1}{n}\sum_{i=1}^{n}(\phi(\bx_i)-G(\bh_t, \epsilon,\bx_i^t))^2
+\frac{\lambda_2}{n_{i}}\sum_{j=1}^{n_{i}}(\nabla\cdot K\cdot\nabla G(\bh_t, \epsilon,\bx_j))^2
+\\\\
\frac{\lambda_3}{n_{l}}\sum_{i=1}^{n_{l}}(\frac{\partial}{\partial x} G(\bh_t, \epsilon,\bx_i^l))^2
+
 \frac{\lambda_4}{n_{r}}\sum_{i=1}^{n_{r}}(\frac{\partial }{\partial x}  G(\bh_t, \epsilon, \bx_i^r))^2
 +
\frac{\lambda_5}{n_{b}}\sum_{i=1}^{n_{b}}(10 \times \frac{\partial }{\partial y} G(\bh_t, \epsilon, \bx_i^b))^2\\
\\
+  \lambda_6[\frac{\Delta x}{3}\big(F(\bh_t, \epsilon,\bx_1)
+4\sum_{i~even \in\{ 2,\cdots,n-1\}}F(\bh_t, \epsilon,\bx_i^t) +2\sum_{i~odd \in\{ 3,\cdots,n-2\}} F(\bh_t, \epsilon,\bx_i^t) +\\\\
F(\bh_t, \epsilon,\bx_n^t) \big)]^2,
\eea
\een
where the weights are re-balanced as follows in each iteration
\ben
\bea{l}
loss_{top}=\frac{1}{n}\sum_{i=1}^{n}(\phi(\bx_i)-G(\bh_t, \epsilon,\bx_i^t))^2,\quad
loss_{eq}=\frac{1}{n_{i}}\sum_{j=1}^{n_{i}}(\nabla\cdot K\cdot\nabla G(\bh_t, \epsilon,\bx_j))^2,\\
loss_{left}=\frac{1}{n_{l}}\sum_{i=1}^{n_{l}}( G_x(\bh_t, \epsilon,\bx_i^l))^2,\quad
loss_{right}=\frac{1}{n_{r}}\sum_{i=1}^{n_{r}}( G_x(\bh_t, \epsilon, \bx_i^r))^2,\\
loss_{bottom}=\frac{1}{n_{b}}\sum_{i=1}^{n_{b}}( G_y(\bh_t, \epsilon, \bx_i^b))^2,\\
loss_{con}=[\frac{\Delta x}{3}\big(F(\bh_t, \epsilon,\bx_1)
+4\sum_{i~even \in\{ 2,\cdots,n-1\}}F(\bh_t, \epsilon,\bx_i^t) +2\sum_{i~odd \in\{ 3,\cdots,n-2\}} F(\bh_t, \epsilon,\bx_i^t)\\
 +F(\bh_t, \epsilon,\bx_n^t) \big)]^2,\\
\overline{loss^{i-1}}=(loss_{top}^{i-1}+loss_{eq}^{i-1}+loss_{right}^{i-1}+loss_{left}^{i-1}+loss_{bottom}^{i-1}+loss_{con}^{i-1})/6,\\
\lambda_1^{i} = loss_{top}^{i-1}/~\overline{loss^{i-1}},\quad
\lambda_2^{i} = loss_{eq}^{i-1}/~\overline{loss^{i-1}},\quad
\lambda_3^{i} = loss_{right}^{i-1}/~\overline{loss^{i-1}},\\
\lambda_4^{i} = loss_{left}^{i-1}/~\overline{loss^{i-1}},\quad
\lambda_5^{i} = loss_{bottom}^{i-1}/~\overline{loss^{i-1}},\quad
\lambda_6^{i} = loss_{con}^{i-1}/~\overline{loss^{i-1}}
\eea
\een

For a given set of randomly chosen top boundary dataset in $\bh_t$: $\bh_t^{(1)}, \cdots, \bh_t^{(m)}$, and the aspect ratio $\epsilon^{(l)}, l=1,\cdots, L$, we define the total loss function as follows
\ben
L(\theta)=\frac{1}{m}\sum_{l=1}^L\sum_{i=1}^m L(\theta, \bh_t^{(i)}, \epsilon^{(l)}).\label{Loss-f}
\een
We remark that the numbers of randomly selected interior and boundary points at each given $\bh_t$ and $\epsilon^{(l)}$ are not the same so that $L(\theta, \bh_t^{(i)}, \epsilon^{(l)})$ can have different number of terms in the sums.
We point it out that choices of activation functions are important to the performance of machine learning model. In this model, we use tanh as the activation function. If one uses DeepONet to solve non-linear equations, it's better off to use smooth activation functions. Our experience with ReLU for this problem is not as good as the one using the tanh function as the activation function.

\subsection{Spectral representation of the boundary}

Alternatively, we represent continuous top boundary $\phi(x)$ using a truncated Sine Fourier series together with a linear interpolation function as follows \cite{habermanAppliedPartialDifferential2013}:
\ben
\phi(x) = \phi(0)+\frac{\phi(L)-\phi(0)}{L}x+\sum_{j=1}^m b_j \sin \frac{j\pi}{L} x,\label{sine}
\een
where $m$ is the number of modes in the spectral expansion and $b_j$ is the jth Sine Fourier coefficient given by
\ben
b_j=\frac{2}{L}\int_{0}^{L}[\phi(x)-\phi(0)-\frac{\phi(L)-\phi(0)}{L}x]\sin(\frac{j\pi}{L} x) dx.
\een
 We represent the top boundary using $m+2$ discrete values $\bh_t=(\phi(0), \phi(L), b_1, \cdots, b_m) \in \bR^{m+2}$, consisting of the Sine Fourier coefficients and the two end point values.
Given the boundary condition at top boundary $\phi(\bx)$, we want to learn a mapping from $(\bh_t, \epsilon)$ to the solution of the steady state governing equation in $\Omega$.

We denote the solution of the boundary value problem in the interior of $\Omega$ by  DeepONet $G(\bh_t, \bx)$ as follows:
\ben
G(\bh_t, \epsilon, \bx)=\sum_{k=1}^p\sum_{i=1}^q c_{i}^k\sigma(\sum_{j=1}^{m+2}\xi_{ij}^k \bh_{t,j}+\xi^k_{i0} \epsilon+\theta_i^k)\sigma(W_k\cdot \bx+\zeta_k)+b_0,\label{G-spectral}
\een
where $c_i^k$, $\xi_{ij}^k, W_k$ are weights and $\theta_i^k, \zeta_k, b_0$ biases.We randomly sample $ n_{l}, n_{r}, n_{b}$ points from the  left, right and bottom boundary respectively, $n_i$ points in the interior of $\Omega$. We divide $[0,1]$ uniformly into $n-1$ intervals, separated by $x_i=(i-1)\Delta x, \Delta x=\frac{1}{n_t-1}, i=1, \cdots, n_t$.
We use the DeepONet to learn the mapping from $(\bh_t, \epsilon)$ to solution $h(\bx,t)$ in $\Omega$.

The cost function in the model for each given top boundary,  the randomly chosen points along the boundary and in the interior, $\{\bx_j^i, i=l,r,b, \bx_j\}$, a well-defined uniformed division of $[0,1]$, $\{x_j, j=1,\cdots, n_t \}$, that defines the top boundary points $ \bx_j^t=(x_j, \phi( x_j))$,  and aspect ratio $\epsilon$  is then defined by
\ben
\bea{l}
L(\theta, \bh_t, \epsilon)= \frac{1}{n_{i}}\sum_{j=1}^{n_{i}}(\nabla \cdot K\cdot \nabla G(\bh_t,\epsilon,\bx_j))^2

+\frac{1}{n_{t}}\sum_{j=1}^{n_{t}}[\phi(\bx_j^t)-G(\bh_t,\epsilon, \bx_j^t)]^2\\\\
+\frac{1}{n_{l}}\sum_{i=1}^{n_{l}}(G_x(\bh_t,\epsilon,\bx_i^l))^2
+
 \frac{1}{n_{r}}\sum_{i=1}^{n_{r}}( G_x(\bh_t, \epsilon,\bx_i^r))^2
 +\frac{1}{n_{b}}\sum_{i=1}^{n_{b}}( G_y(\bh_t, \epsilon,\bx_i^b))^2
 \\\\
+ [\frac{\Delta x}{3}\big(F(\bx_1, \epsilon)
+4\sum_{i~even \in\{ 2,\cdots,n-1\}}F(\bx_i^t, \epsilon) +2\sum_{i~odd \in\{ 3,\cdots,n-2\}} F(\bx_i^t, \epsilon) +F(\bx_n^t, \epsilon) \big)]^2.
\eea
\een
 In the practical implementation, we once again re-balance the ``local loss" as alluded to earlier.

For the bounded Toth basin with two vertical, lateral boundaries, we can rescale or transform  the bounded, arbitrary physical domain into a rectangular domain and then solve the equation in the rectangular domain. We call this the domain mapping approach.

 \subsection{Domain mapping}

We present yet another alternative approach to establish the mapping from the top boundary to the solution in a Toth basin using a nonlinear domain mapping. We assume the top boundary is given by $y=\phi(x)>0$ and the bottom one by $y=0$ for $x\in [0,L]$. We introduce a change of variable from $(x,y)$ to $(x,z)$ as follows
\ben
x=x, \quad z=\frac{y}{\phi(x)}, \quad y\in [0, \phi(x)].
\een
The 2D gradient operator in the new coordinate is given by
\ben
\nabla^*=(\frac{\partial}{\partial x}, \frac{\partial }{\partial z})=(\frac{\partial}{\partial x}+\frac{y}{\phi}\phi_x \frac{\partial}{\partial y},  \phi\frac{\partial }{\partial y}).
\een
Or equivalently,
\ben
\bea{l}
\nabla=(\frac{\partial}{\partial x}, \frac{\partial }{\partial y})=(\frac{\partial}{\partial x}-\frac{y}{\phi^2}\phi_x \frac{\partial }{\partial z}, \frac{1}{\phi}\frac{\partial }{\partial z})\\
=
\left(
\bea{ll}
1 & -\frac{y}{\phi^2}\phi_x\\
0 & \frac{1}{\phi(x)}
\eea\right)\cdot \nabla^*.
\eea
\een
We denote
\ben
D=\left(
\bea{ll}
1 & -\frac{y}{\phi^2}\phi_x\\
0 & \frac{1}{\phi(x)}
\eea\right)=\left(
\bea{ll}
1 & -\frac{z}{\phi}\phi_x\\
0 & \frac{1}{\phi(x)}
\eea\right).
\een
The Laplace equation is rewritten into
\ben
(D\cdot \nabla^*) \cdot K \cdot (D \cdot \nabla^*) h=0.
\een
The boundary conditions of $h$ is given by
\ben
\left\{
\bea{l}
\bn \cdot K \cdot \nabla h|_{\Gamma_{l, r, b}}=\bn \cdot K\cdot( D\cdot \nabla^*) h|_{\Gamma_{l, r, b}}=0,
 \\
h(x,1)|_{\Gamma_{top}}=\phi(x)\quad  \hbox{(Dirichlet)},\\
 \bn \cdot K\cdot( D\cdot \nabla^*) h|_{\Gamma_{top}}=-\gamma(h-\phi) \quad \hbox{(Robin)},  \quad x\in [0,1].
 \eea\right.
\een
In this study, we limit ourselves to
\ben
K=Diag(K_{11}, K_{22}).
\een
Then,
\bena
\bn\cdot K \cdot (D \cdot \nabla^*) h|_{\Gamma_{l, r}}= K_{11} \frac{\partial h}{\partial x}|_{\Gamma_{l, r}}=0, \\
\bn\cdot K \cdot (D \cdot \nabla^*) h|_{\Gamma_{b}}=\frac{K_{22}}{\phi}\frac{\partial h}{\partial z}|_{\Gamma_{b}}=0.
\eena
These imply
\ben
\frac{\partial h}{\partial x}|_{\Gamma_{l, r}}=0, \quad \frac{\partial h}{\partial z}|_{\Gamma_{b}}=0,
\een
where the bottom boundary is assumed flat.

The steady state governing equation  without a source is given by
\bena
(D\cdot \nabla^*) \cdot K \cdot (D \cdot \nabla^*) h\\\\
=K_{11}[ \parls{h}{x}-\parl{}{x}(\frac{z\phi_x}{\phi}\parl{h}{z} ) -\frac{z\phi_x}{\phi} \parlm{h}{x}{z}+\frac{z\phi_x}{\phi} \parl{}{z} ( \frac{z\phi_x}{\phi} \parl{h}{z})]+\frac{K_{22}}{\phi} \parl{}{z} (\frac{1}{\phi} \parl{h}{z})=0.
\eena
The consistency condition becomes
\ben
\int_{z=1}[\bn^{(t)} \cdot K \cdot (D \cdot \nabla^*) h]~ dx=\int_{z=1}[K_{22}\frac{h_z}{\phi} -K_{11}\phi_x(h_x-\frac{\phi_x}{\phi} h_z)]dx=0,
\een
where $\bn^{(t)}=(-\phi_x, 1)$.
We represent the top boundary using $m+2$ discrete values $\bh_t=(\phi(a), \phi(b), b_1, \cdots, b_m)$ from the truncated Sine Fourier series approximation.
Given the boundary condition at the top boundary $z=1$, we want to learn a mapping from $\bh_t$ to the solution of the steady state governing equation in $\Omega$.

We denote the solution in the interior of $\Omega$ by DeepONet $G(\bh_t, \bx)$  defined in \eqref{G-spectral}.
The loss function  for each given top boundary $\bh_t$, aspect ratio $\epsilon$, the randomly chosen points along the boundary and in the interior, $\{\bx_j^i, i=l,r,b,  \bx_k\}$, and a well-defined uniform division of $[0,1]$, $\{x_j, \}$, that defines the top boundary points $ \bx_j^t=(x_j, 1)$,  is then defined by
\ben
\bea{l}
L(\theta, \bh_t, \epsilon)= \frac{1}{n_{i}}\sum_{j=1}^{n_{i}}[(D\cdot \nabla^*)\cdot  K\cdot (D\cdot \nabla^*) G(\bh_t, \epsilon,\bx_j^t)]^2
+\frac{1}{n_{t}}\sum_{j=1}^{n_{t}}[\phi(\bx_j^t)-G(\bh_t, \epsilon, \bx_j^t)]^2\\\\
+\frac{1}{n_{l}}\sum_{i=1}^{n_{l}}[K_{11}G_x(\bh_t, \epsilon,\bx_i^l)]^2
+ \frac{1}{n_{r}}\sum_{i=1}^{n_{r}}[ K_{11}G_x(\bh_t, \epsilon, \bx_i^r)]^2\\\\
+\frac{1}{n_{b}}\sum_{i=1}^{n_{b}}[\bn \cdot K\cdot( D \cdot \nabla^*) G(\bh_t, \epsilon, \bx_i^b)]^2
+ [\frac{\Delta x}{3}(F(\bh_t, \epsilon,\bx_1^t)+\\\\
4\sum_{i~even \in\{2,\cdots, n_t-1\}} F(\bh_t, \epsilon,\bx_i^t) +2\sum_{i~odd \in\{3,\cdots, n_t-2\}} F(\bh_t, \epsilon,\bx_i^t)+F(\bh_t, \epsilon,\bx_{n_t}))]^2,
\eea
\een
where
\ben
F(\bx, \epsilon)=K_{22}\frac{G_z}{\phi} -K_{11}\phi_x(G_x-\frac{\phi_x}{\phi} G_z).
\een

 In the practical implementation, we adopt a re-balanced or modified loss function, in which we add a weight to each term in the loss.  The modified loss function is given by
\ben
\bea{l}
L(\theta, \bh_t, \epsilon)=\frac{\lambda_1}{n_{t}}\sum_{j=1}^{n_{t}}[\phi(\bx_j^t)-G(\bh_t, \epsilon,\bx_j^t)]^2

+\frac{\lambda_2}{n_{i}}\sum_{j=1}^{n_{i}}[(D \cdot \nabla^*) \cdot K\cdot ( D \cdot \nabla^*) G(\bh_t, \epsilon,\bx_j)]^2
\\\\
+\frac{\lambda_3}{n_{l}}\sum_{i=1}^{n_{l}}[G_x(\bh_t, \epsilon,\bx_i^l)]^2
+
 \frac{\lambda_4}{n_{r}}\sum_{i=1}^{n_{r}}[ G_x(\bh_t, \epsilon, \bx_i^r)]^2
+\frac{\lambda_5}{n_{b}}\sum_{i=1}^{n_{b}}[10 \times \bn \cdot K\cdot( D \cdot \nabla^*) G(\bh_t, \epsilon, \bx_i^b)]^2
\\\\
+ \lambda_6[\frac{\Delta x}{3}(F(\bx_1^t, \epsilon)+4\sum_{i~even \in\{2,\cdots, n_t-1\}} F(\bx_i^t, \epsilon) +2\sum_{i~odd \in\{3,\cdots, n_t-2\}} F(\bx_i^t, \epsilon)+F(\bx_{n_t}, \epsilon))]^2,
\eea
\een
where the weights are re-balanced as follows in each iteration
\ben
\bea{l}
loss_{top}=\frac{1}{n_t}\sum_{i=1}^{n_t}(\phi(\bx_i^t)-G(\bh_t,\bx_i^t))^2,\quad
loss_{eq}=\frac{1}{n_{i}}\sum_{j=1}^{n_{i}}((D\cdot \nabla^*)\cdot K\cdot(D\cdot \nabla^*) G(\bh_t,\bx_j))^2,\\
loss_{left}=\frac{1}{n_{l}}\sum_{i=1}^{n_{l}}( G_x(\bh_t,\bx_i^l))^2,\quad
loss_{right}=\frac{1}{n_{r}}\sum_{i=1}^{n_{r}}( G_x(\bh_t, \bx_i^r))^2,\\
loss_{bottom}=\frac{1}{n_{b}}\sum_{i=1}^{n_{b}}(\frac{K_{22}}{\phi}G_z(\bh_t, \bx_i^b))^2,\\
loss_{con}=[\frac{\Delta x}{3}(F(\bx_1^t)+4\sum_{i~even \in\{2,\cdots, n_t-1\}} F(\bx_i^t) +2\sum_{i~odd \in\{3,\cdots, n_t-2\}} F(\bx_i^t)+F(\bx_{n_t}))]^2\\

\overline{loss^{i-1}}=(loss_{top}^{i-1}+loss_{eq}^{i-1}+loss_{right}^{i-1}+loss_{left}^{i-1}+loss_{bottom}^{i-1}+loss_{con}^{i-1})/6,\\
\lambda_1^{i} = loss_{top}^{i-1}/~\overline{loss^{i-1}},\quad
\lambda_2^{i} = loss_{eq}^{i-1}/~\overline{loss^{i-1}},\quad
\lambda_3^{i} = loss_{right}^{i-1}/~\overline{loss^{i-1}},\\
\lambda_4^{i} = loss_{left}^{i-1}/~\overline{loss^{i-1}},\quad
\lambda_5^{i} = loss_{bottom}^{i-1}/~\overline{loss^{i-1}},\quad
\lambda_6^{i} = loss_{con}^{i-1}/~\overline{loss^{i-1}}.
\eea
\een
The total loss is defined in \eqref{Loss-f} for a given set of top boundaries.
In the rescaled domain, the variable coefficient Poisson equation is solved in a rectangular domain using PIML.

\subsection{Arbitrary bottom boundary}

When the bottom boundary is varying in space as well, we denote it as $\phi_2(x)$. We rescale the physical domain in the y direction as follows
\ben
z=\frac{y-\phi_2(x)}{\phi(x)-\phi_2(x)},\quad y\in [\phi_2(x), \phi(x)].
\een
This mapping transforms the Toth basin into a rectangular domain in a new coordinate. The gradient operator is transformed as follows
\ben
\nabla=(\frac{\partial }{\partial x}-(\frac{z}{\phi-\phi_2} (\phi-\phi_2)_x+\frac{\phi_{2,x}}{\phi-\phi_2})\frac{\partial }{\partial z}, \frac{1}{\phi-\phi_2}\frac{\partial }{\partial z})=D\cdot \nabla^*,
\een
where
\ben
D=\left (
\bea{lr}
1 & -(\frac{z}{\phi-\phi_2}(\phi-\phi_2)_x+\frac{\phi_{2,x}}{\phi-\phi_2})\\
0 & \frac{1}{\phi-\phi_2}
\eea
\right).
\een

The Laplace equation is rewritten into a variable coefficient one as follows
\ben
(D\cdot \nabla^*) \cdot K \cdot (D \cdot \nabla^*) h=0,
\een
where
\ben
\bea{l}
(D\cdot \nabla^*) \cdot K \cdot (D \cdot \nabla^*) h=K_{11}[\frac{\partial^2}{\partial x^2}-\frac{\partial}{\partial x}((\frac{z}{\phi-\phi_2} (\phi-\phi_2)_x+\frac{\phi_{2,x}}{\phi-\phi_2})\frac{\partial }{\partial z})-(\frac{z}{\phi-\phi_2} (\phi-\phi_2)_x\\
\\
+\frac{\phi_{2,x}}{\phi-\phi_2})\frac{\partial^2 }{\partial z \partial x}+(\frac{z}{\phi-\phi_2} (\phi-\phi_2)_x+\frac{\phi_{2,x}}{\phi-\phi_2})\frac{\partial }{\partial z}((\frac{z}{\phi-\phi_2} (\phi-\phi_2)_x+\frac{\phi_{2,x}}{\phi-\phi_2})\frac{\partial }{\partial z})]+
K_{22}\frac{1}{(\phi-\phi_2)^2}\frac{\partial^2}{\partial z^2}=0.
\eea
\een
The boundary conditions of $h$ are given by
\bena
\bn \cdot K \cdot \nabla h|_{\Gamma_{l, r, b}}=\bn \cdot K \cdot(D\cdot \nabla^*) h|_{\Gamma_{l, r, b}}=0,
\\
h(x,1)_{\Gamma_{top}}=\phi(x) \quad  \hbox{(Dirichlet)}, \quad \bn \cdot K \cdot(D\cdot \nabla^*) h|_{\Gamma_{top}}=-\gamma(h-\phi)\quad  \hbox{(Robin)}, \quad x\in [0,1],
\eena
where $\bn=(\pm 1, 0)$ are the unit external normal of the lateral surfaces, $\bn=\frac{1}{\sqrt{1+\phi_{2,x}^2}}( -\phi_{2,x},1)$ is the external unit normal of the bottom surface.
Namely,
\ben
\frac{\partial h}{\partial x}|_{left, right}=0, \quad \bn\cdot K \cdot(D \cdot \nabla^*) h|_{bottom}=0.
\een
The consistency condition for the boundary conditions is
\ben
\int_{z=1} \bn^{(t)} \cdot K\cdot (D\cdot \nabla^*) h dx=\int_{z=1} [\frac{K_{22}}{\phi-\phi_2}h_z-K_{11}\phi_x
(h_x -\frac{\phi_{x}}{\phi-\phi_2} h_z)]dx=0.
\een
We define
\ben
F(\bx, \epsilon)=[\frac{K_{22}}{\phi-\phi_2}h_z-K_{11}\phi_x
(h_x -\frac{\phi_{x}}{\phi-\phi_2} h_z)].
\een
Analogous to the treatment of the top boundary, we expand $\phi_2$ in its truncated Fourier Sine series
\ben
\phi_2=\phi_2(a)+\frac{\phi_2(b)-\phi_2(a)}{L}(x-a)+\sum_{i=1}^m c_i \sin (i\pi \frac{x}{L}).
\een
We denote
\ben
\bh_t=(\phi(a), \phi(b), b_1, \cdots, b_n, \phi_2(a), \phi_2(b), c_1, \cdots, c_m).
\een
The DeepONet is defined by the following:
\ben
G(\bh_t, \epsilon, \bx)=\sum_{k=1}^p\sum_{i=1}^q c_{i}^k\sigma(\sum_{j=1}^{n+m+4} \xi_{ij}^k \bh_{t,j}+\xi_{10}^k \epsilon+\theta_i^k)\sigma(W_k\cdot \bx+\zeta_k)+b_0.
\een
 The loss function  is given by
 \ben
\bea{l}
L(\theta, \bh_t, \epsilon)= \frac{1}{n_{i}}\sum_{j=1}^{n_{i}}[(D\cdot \nabla^*)\cdot  K\cdot (D\cdot \nabla^*) G(\bh_t, \epsilon,\bx_j^t)]^2

+\frac{1}{n_{t}}\sum_{j=1}^{n_{t}}[\phi(\bx_j^t)-G(\bh_t, \epsilon, \bx_j^t)]^2\\\\

+\frac{1}{n_{l}}\sum_{i=1}^{n_{l}}[ G_x(\bh_t, \epsilon,\bx_i^l)]^2

+
 \frac{1}{n_{r}}\sum_{i=1}^{n_{r}}[  G_x(\bh_t, \epsilon, \bx_i^r)]^2
+\frac{1}{n_{b}}\sum_{i=1}^{n_{b}}[\frac{1}{\phi} G_z(\bh_t, \epsilon, \bx_i^b)]^2
\\\\
+ [\frac{\Delta x}{3}(F(\bx_1^t, \epsilon)+4\sum_{i~even \in\{2,\cdots, n_t-1\}} F(\bx_i^t, \epsilon) +2\sum_{i~odd \in\{3,\cdots, n_t-2\}} F(\bx_i^t, \epsilon)+F(\bx_{n_t}, \epsilon))]^2.
\eea
\een
where $\bx_j$ are the interior points and $\bx_j^k$ are boundary points at the top, left, right, and bottom boundary, respectively. The total loss is given by \eqref{Loss-f} when a set of top and bottom boundaries are given. In practice, the modified loss function is adopted analogous to what we alluded to earlier.

\section{Results and discussion}

We present the numerical results in two scenarios. In the first scenario, we learn the mapping from an arbitrary surface topography to the solution in the basin while the bottom boundary is assumed flat. In the second, we allow the bottom boundary to be an arbitrary shape as well. We have implemented all the methods using PyTorch. For simplicity, we present the results obtained using the spectral representation and the domain mapping method only in the following. We remark that different surface representations produce the same numerical result.

\subsection{Results with an arbitrary top boundary}

We first present the results obtained using the spectral representation method.

\subsubsection{Sampling of the top boundary representation}
The top boundary is defined by \eqref{sine} with coefficients or parameters in $\bh_t$. The sampling of $\bh_t=(\phi(a), \phi(b), b_1, \cdots, b_m)$ is carried out as follows
\begin{itemize}
\item We sample $\phi(a)$ uniformly from $[0.7,0.8]$ and  $\phi(b)$ uniformly from $[\phi(a)-0.2,\phi(a)+0.2]$ to ensure that fluctuations of the boundary function are reasonable geographically.
\item  For $h_0(x)=\sum^{8}_{j=1}b_j\sin(j\pi x)$ with $m=8$, we sample $b_1, \cdots b_8$ uniformly from $[-1,1]$, respectively.
\item We calculate $h_{max}=\max_{x\in [0,1]} \phi(x), h_{min}=\min_{x\in [0,1]} \phi (x)$, and $h_d=h_{max}-h_{min}$. We sample $\lambda \in [0,0.2]$ and then update coefficients $b_j:=\lambda b_j/h_d, j=1,\cdots, 8.$
    \item Then, the  top boundary surface is well-represented by vector $\bh_t=(\phi(a), \phi(b), b_1, \cdots, b_8)$.
    \item For illustration purposes, we set $\epsilon=0.01$ throughout the paper.
\end{itemize}
This sampling method makes sure the top boundary fluctuates in $0.5<\phi(x)\leq 1$. The larger fluctuation can be done, but it may not be necessary for the realistic geography.

\subsubsection{The dataset}

The Loss function is defined by summing up all the squared residues of the equation and the boundary conditions as well as a consistency condition that depends on $\bh_t$. We denote the input of the neural network in the loss function as follows:
\ben
\bz=(\bh_t, \bx),
\een
where $\bx$ is a long vector containing randomly chosen points from the boundaries and the interior of the basin underneath the top boundary represented by $\bh_t$ which are chosen after the top boundary  is specified.
We sample I number of representing vectors of top boundaries in $\bh_t^{i}, i=1, \cdots, I$. For each $1\leq i\leq I$, we have well-defined top boundary $y=\phi_i(x)$. For the ith top boundary, we randomly choose $L_i$ data points $\bx_j^{l,i}, j=1,\cdots, L_i$ on the left boundary, $R_i$ points $\bx_{j}^{r,i}, j=1,\cdots, R_i$ on the right boundary, $B_i$ points $\bx_j^{b,i}, j=1,\cdots, B_i$ on the bottom,  and $J_i$ points in the interior $\bx_j^i \in [0, 1]\times [0, \phi(\bx)], j=1,\cdots, J_i$, $M$ uniform points  $x_j, j=1,\cdots, M$ in $[0,1]$.

We divide the dataset into the training and test sets by randomly dividing $\{1, \cdots, I\}$ into two subsets $I_{train}$ and $I_{test}$.
We generate 140 top boundary topographies using the spectral representation.   For 100 boundary topographies, we sample 14000 points randomly, including 10000 interior points and 4000 boundary points.  For the rest 40 boundary topographies, we put them in the test set.  Finally, we choose $M=101$ points uniformly in $[0,1]$ to calculate the consistency condition included in the loss function.

In the  DeepONet, the width of branch and trunk net is 200, the depth of the branch net is 4, and the depth of the trunk net is 3. We use the Adam algorithm for the first 1000 epoch optimization step with learning rate $10^{-4}$ and weight decay $10^{-7}$. For the remaining epoch, we use the LBFGS algorithm with learning rate $0.1$. The stopping criterion of the LBFGS is for both training and testing processes. The typical loss in each weighted component in the modified loss function at the end of the machine learning process is summarized  in Figure \ref{Loss-summary}.
For the parameters in the model, we use characteristic length scales $L_{y}$=1000 m and $L_{x}$=10000 m, which lead to $k_{11}$=0.01 and $k_{22}$=1 (i.e., aspect ratio $\epsilon=0.01$). All model parameters are summarized in Table \ref{table1}.

\begin{table}[ht]
\caption{Typical loss values at the end of training and testing, respectively}\label{Loss-summary}
\vskip 12 pt
\centering
\begin{tabular}{|c|c|c|}
\hline
\hline
Loss terms& Training loss& Test loss\\
\hline
$Loss_{eq}$& $4.82\times 10^{-4}$& $2.35\times 10^{-3}$\\
\hline
$Loss_l$& $4.81\times 10^{-7}$ & $7.57\times 10^{-6}$\\
\hline
$Loss_r$ & $5.6\times 10^{-7}$ & $1.71\times 10^{-5}$\\
\hline
$Loss_b$ & $5.07\times 10^{-5}$ & $2.18\times 10^{-5}$\\
\hline
$Loss_t$ & $4.58\times 10^{-4}$& $7.77\times 10^{-3}$\\
\hline
$Loss_{con}$ & $1.69\times 10^{-5}$ & $3.38\times 10^{-3}$\\
\hline
\end{tabular}
\end{table}

\begin{table}[ht]
\caption{Model, neural network and optimization algorithm parameters }\label{table1}
\vskip 12 pt
\centering
\begin{tabular}{|c|c|}
\hline
\hline
Parameter& Value\\
\hline
Width of trunk and branch net& 200\\
\hline
Depth of branch net & 4\\
\hline
Depth of trunk net & 3\\
\hline
Weight decay & $10^{-7}$\\
\hline
Learning rate for the first 1000 epoch & $10^{-4}$\\
\hline
Learning rate for the remaining epoch & 0.1\\
\hline
$L_y$ & 1000 m\\
\hline
$L_x$ & 10000 m\\
\hline
$k_{11}$ & 0.01\\
\hline
$k_{22}$ & 1\\
\hline
\end{tabular}
\end{table}

\subsubsection{Benchmark with the Toth water table solution and the numerical solution obtained using COMSOL}

\begin{figure}[H]
\centering
\subfigure[The relative error between the solution obtained from the DeepONet and the Toth's solution. $RMSE=6.2\times 10^{-3}$.]
{\includegraphics[width=0.9\textwidth]{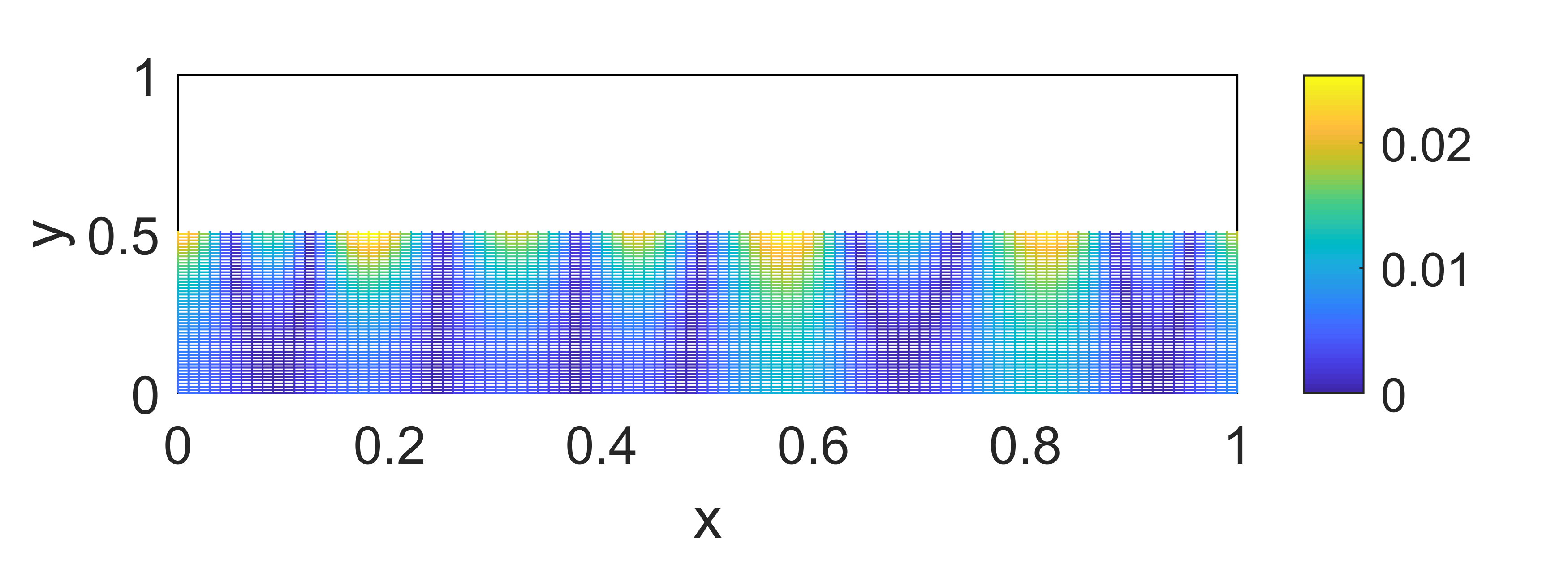}}
\subfigure[The relative error between the solution obtained from the DeepONet and the numerical one from COMSOL. Here $RMSE=2.05\times 10^{-2}$.]
{\includegraphics[width=0.9\textwidth]{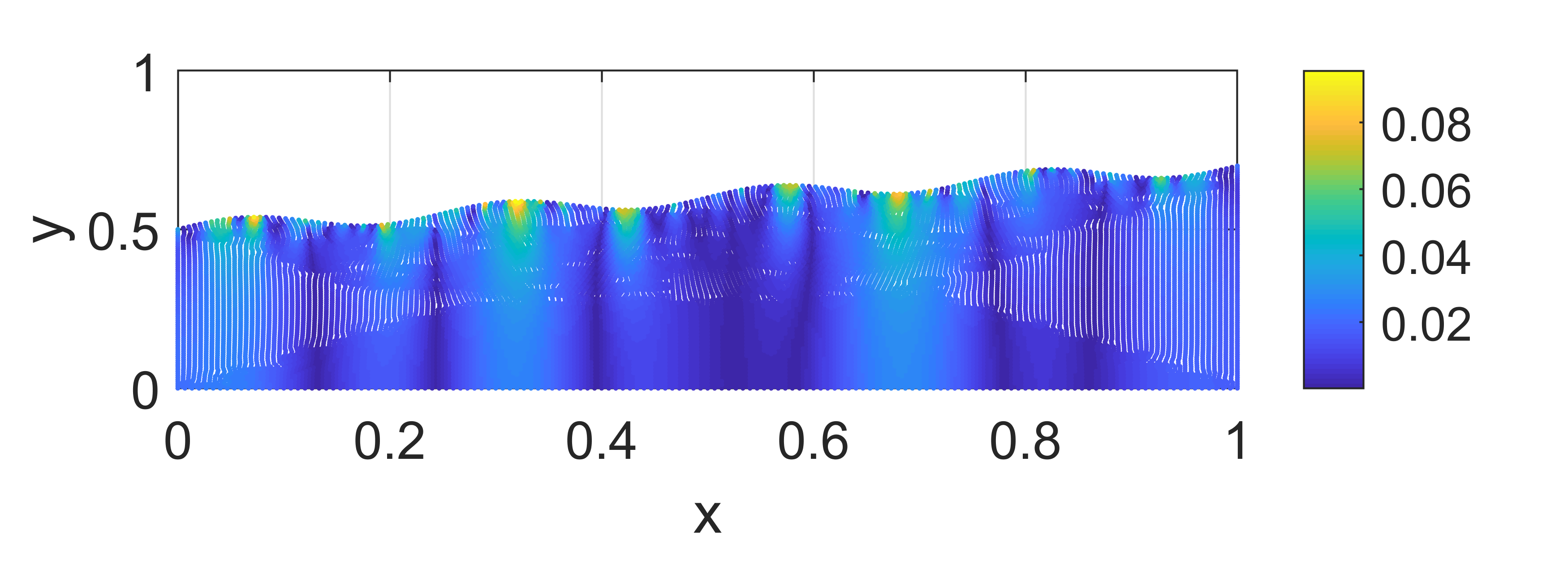}}
\caption{The  top boundary is given by $y_t=y_0+c^{\prime}x+a^{\prime}sinb^{\prime}x$, which $\tan\alpha=c^{\prime}, a/\cos\alpha=a^{\prime}, b/\cos\alpha=b^{\prime}$, $y_0=0.5, c^{\prime}=0.02, a=0.025$.}\label{comp-fig}
\end{figure}

We compare the solution obtained from the DeepONet mapping and the Toth's water table solution and the numerical solution obtained from COMSOL for a given top surface. The results are summarized in Figure \ref{comp-fig}. The relative mean square errors (RMSEs) are in the order of $10^{-2}$. Given the loss in machine-learning is in the order of $10^{-3}$,  the Toth's solution is asymptotic, and the numerical solution is approximate, the RMSEs are consistent with the errors one expects from the PIML approach.

\subsubsection{Results obtained using the spectral representation }

\begin{figure}[H]
\centering
\subfigure[$\bh_t=(0.75, 0.95, 0, 0, 0, 0, 0, 0, 0, 0.015).$]{
			\includegraphics[width=1\textwidth]{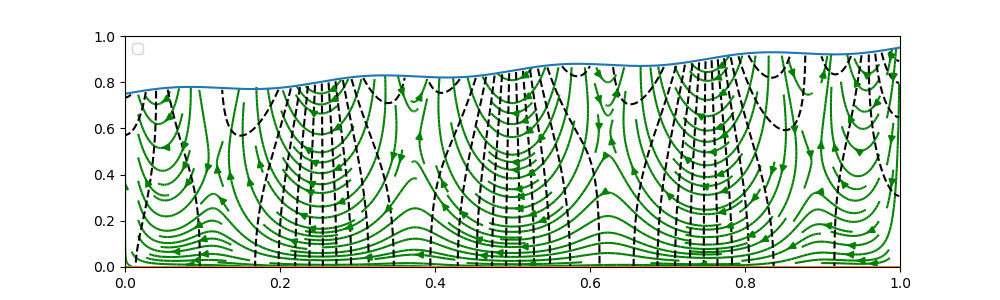}
}\\
\subfigure[$\bh_t= (0.75, 0.95, 0, 0, 0, 0, 0, 0, 0, 0.045).$]{
			\includegraphics[width=1\textwidth]{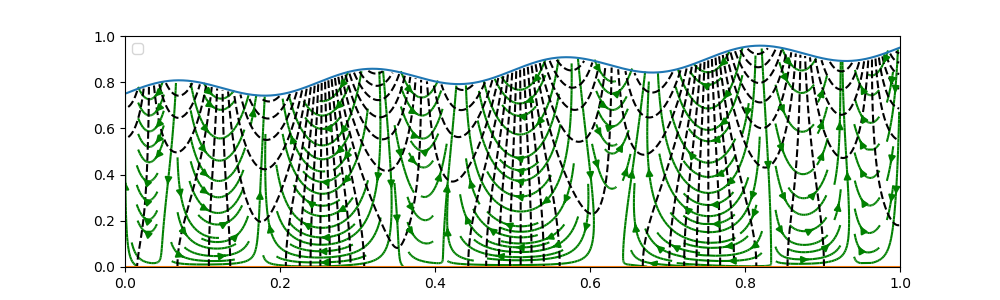}
}
%
\caption{Steady flow patterns in streamlines and potential distributions in the Toth basin with a sloped and localized spatial variations at the  top boundary represented by a spectral representation.  (a). The slow varying surface topography promotes long distance transport at the bottom of the basin due to the slope. (b).
 When the localized spatial  variations in the top are enhanced, the number of  compartmentalized circulations increases
and flows are blocked to compartmentalized circulations, cutting off the long distance transport. }
\label{12}
\end{figure}
\begin{figure}[H]
\centering
\subfigure[$\bh_t=(0.75, 0.95, 0, 0, 0, 0, 0, 0, 0, 0.015).$]{
			\includegraphics[width=0.8\textwidth]{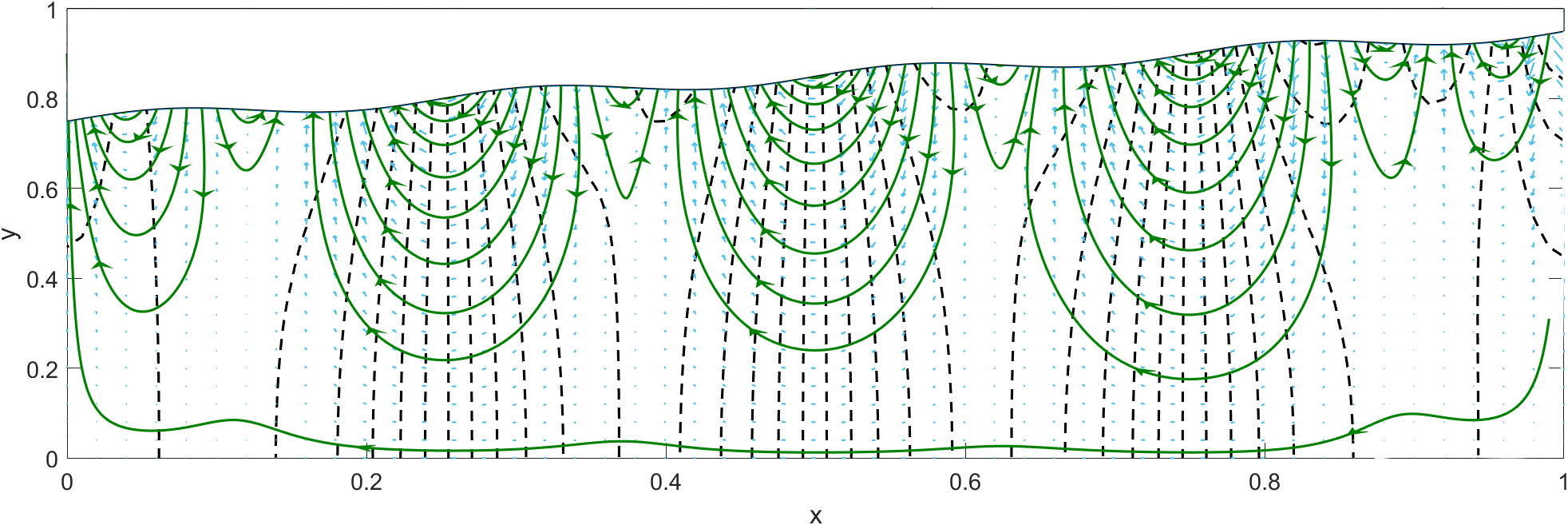}
}
\subfigure[$\bh_t= (0.75, 0.95, 0, 0, 0, 0, 0, 0, 0, 0.045).$]{
			\includegraphics[width=0.8\textwidth]{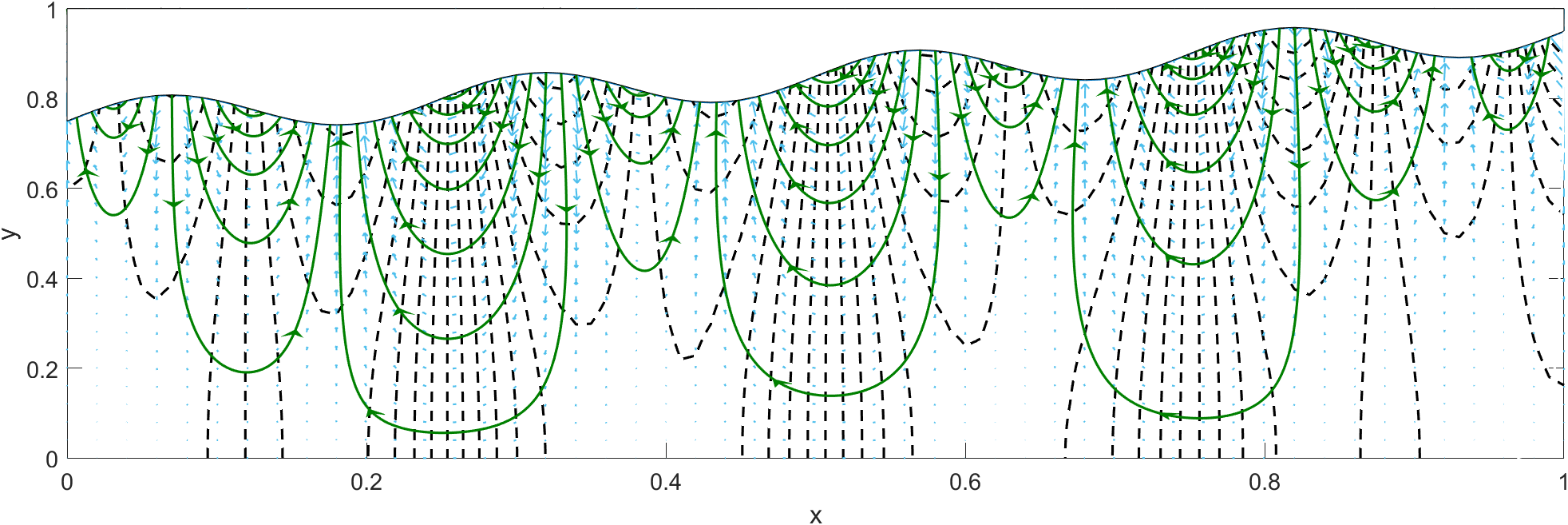}
}
\caption{Replot of the steady solutions in Figure \ref{12} with continuous streamlines and potential distributions. These plots reconstruct some continuous long streamlines partially shown in Figure \ref{12}.  }
\label{12-2}
\end{figure}

After learning the mapping represented by the DeepONet, we present several representative results obtained using the mapping to show the  solution of the steady state groundwater flow equation in the Toth basin.  Figure \ref{12} depicts the flow field and the water potential distribution in a Toth basin with  sloped top surface topographies of localized variations. There are two factors  in the top boundary that impact on flow patterns in the Toth basin: one is the average slope and the other is the localized variation of the surface. A large average slope tends to promote long distance transport of the flow at the bottom of the basin in addition to the compartmentalized or localized circulatory flow patterns near the top surface. When localized variations in the top surface are large, the long distance transport near the bottom tends to be blocked by intruding localized circulations penetrated down from the top. Figure \ref{12} shows a typical steady state flow pattern due to the surface  topography with a fixed average slope and varying localized surface variations in space. The two topographical features identified and their influence to steady state flow patterns are shown visibly. To render a better graphical resolution for some long streamlines, we use an image reconstruction method to reconstruct some of the continuous long streamlines that are not well-shown in Figure \ref{12}. The results are depicted in Figure \ref{12-2}.

\begin{figure}[H]\centering
 \subfigure[$\bh_t= (0.75, 0.85, 0, 0, 0, 0, 0, 0, 0, 0.015).$]{
			\includegraphics[width=1\textwidth]{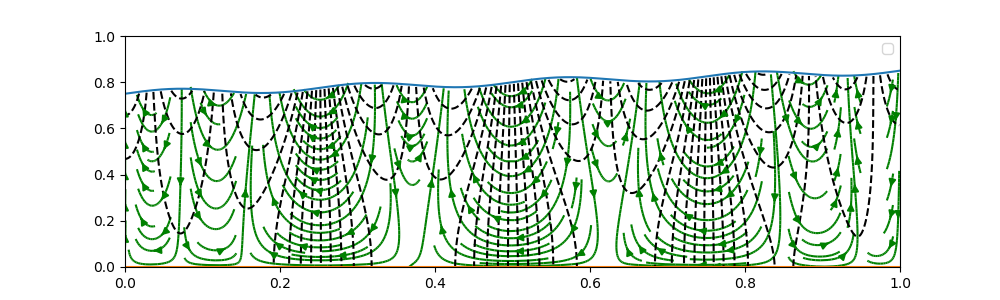}
}\\
\subfigure[$\bh_t= (0.75, 0.75, 0, 0, 0, 0, 0, 0, 0, 0.015).$]{
			\includegraphics[width=1\textwidth]{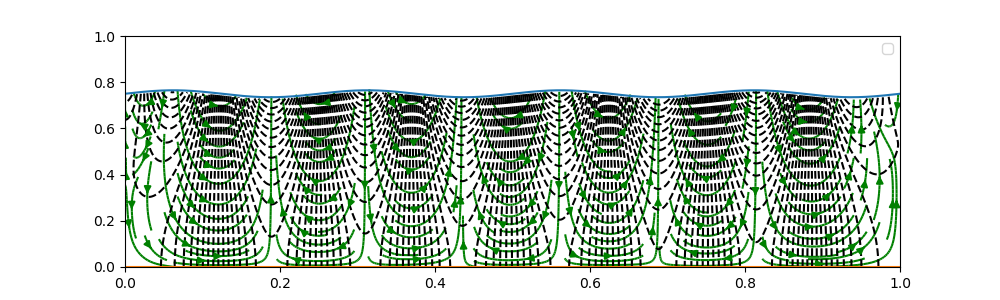}
}
\caption{Steady flow patterns with top surface of small slopes. Small slopes in the surface topography with small surface fluctuations lead to compartmentalized circulations. A top boundary with a zero average slope  separates all flows into compartmentalized circulations of nearly equal width. }
\label{14}
\end{figure}

Consistent with  Toth's results \cite{Toth1963}, our solutions also show that a small average slope in the topography and small fluctuation in spatial variations promotes compartmentalized circulations. Figure \ref{14} shows two cases of flow fields with small average slopes of the top surface. A top surface with a zero average slope and some spatial variations creates several fully compartmentalized flow patterns correlated with the wave form of the top boundary.

\begin{figure}[H]
\centering
 \subfigure[]{
			\includegraphics[width=1\textwidth]{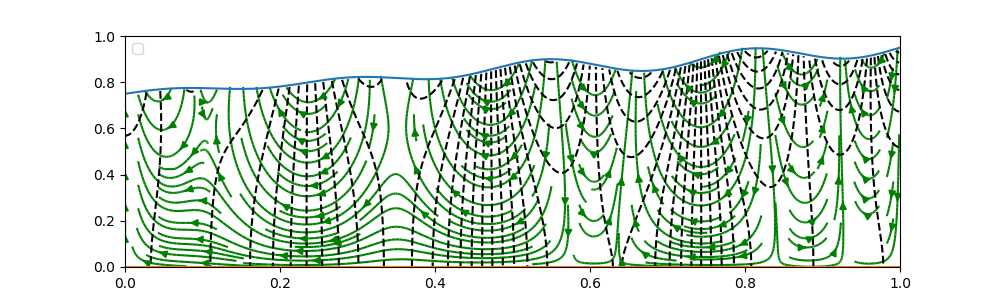}
}\\
\subfigure[]{
			\includegraphics[width=1\textwidth]{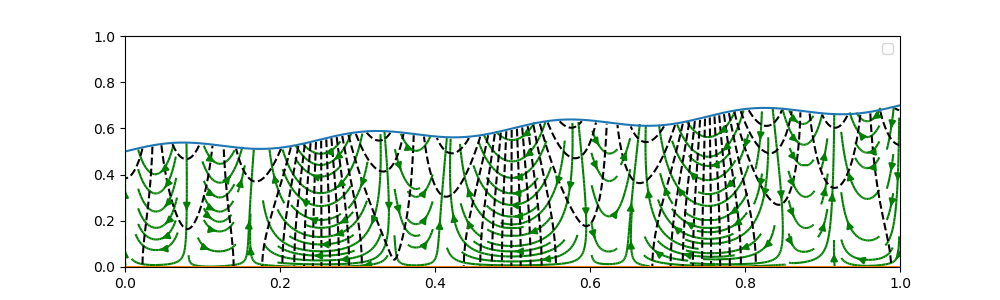}
}
\caption{(a). The curvature is smaller on the left than that on the right.
Parameter values are $\bh_t=[0.75, 0.95, 3.619\times10^{-3}, -1.561\times10^{-16}, -4.418\times10^{-3}, 4.224\times10^{-16}, 5.858\times10^{-3}, 1.472\times10^{-16}, -1.526\times10^{-2}, 2.25\times10^{-2}]$. (b). The curvature enhances on the left in this plot. Parameter values are $\bh_t=[0.5, 0.7, -1.276\times10^{-6}, 2.679\times10^{-6}, -4.382\times10^{-6}, 6.69\times10^{-6}, -1.028\times10^{-5}, 1.717\times10^{-5}, -3.732\times10^{-5}, 0.025]$.}
\label{19}
\end{figure}

We observe that local curvatures in the top surface affect flow patterns in the Toth basin as well. A larger local curvature tends to create more localized flow patterns while the smaller one promotes more global flow patterns in the bottom of the basin. Figure  {\ref{19}} depicts an example where the magnitude of the curvature of the top surface at the left is smaller than that on the right. As the result, the flow patten is more localized at the right than that on the left.  We note that it is the overall slope of the surface topography that dominates the overall flow pattern, while the local curvature makes  the flow pattern more localized (or circulatory). There apparently exists a competition between the local curvature effect and the overall slope of the top boundary. The flow pattern in the classical Toth water table resembles the flat topographical surface shown in Figure \ref{19}-b since the Toth's solution is an asymptotic  one over a near flat top boundary. The current study indeed extends the asymptotic analyses in \cite{Toth1963} to a truly nonlinear topography of potentially large spatial variations. With the boundary-to-solution mapping given by the DeepONet, we can literally calculate any solution pattern as we need as long as the top surface topography is given.


\subsubsection{Results obtained by the domain mapping method}

\begin{figure}[H]
\centering
 \subfigure[$\bh_t=(0.75, 0.95, 0, 0, 0, 0, 0, 0, 0, 0.015).$]{
			\includegraphics[width=1\textwidth]{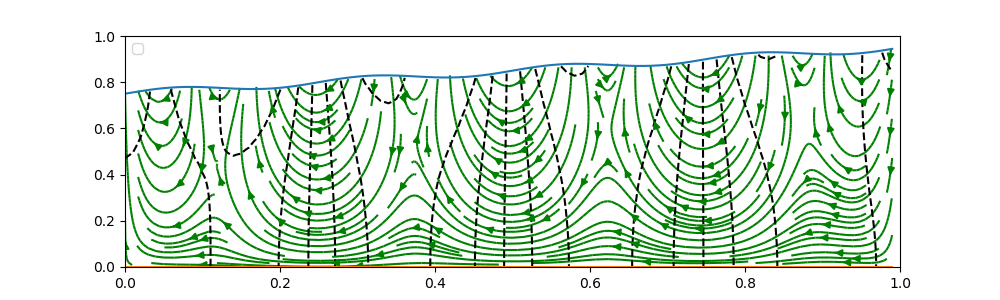}
}\\
 \subfigure[$\bh_t=(0.75, 0.75, 0, 0, 0, 0, 0, 0, 0, 0.015).$]{
			\includegraphics[width=1\textwidth]{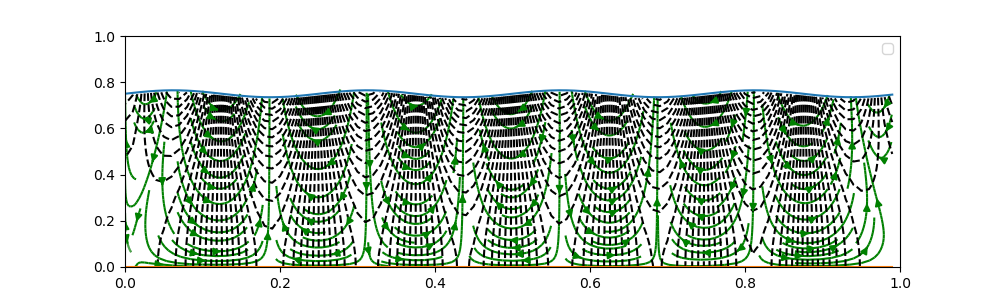}
}
\caption{Steady state solutions obtained using the domain mapping method. (a).  This is identical to  the one with the same parameters  in Figure {\ref{12}}-a. (b).   This is  identical to  the one with the same parameters in Figure  {\ref{14}-b}. Thus, the two methods produce the same results. }
\label{114}
\end{figure}

Here we report the results obtained using the domain mapping method on the same two top boundaries as Figure \ref{12}-a and \ref{14}-b and show that the results are the same numerically. Figure \ref{114} shows two calculated flow fields using the DeepONet obtained from the domain mapping method using the same top surface representations as those used in Figure \ref{12}-a and \ref{14}-b. The results look identical.  Thus, either  method can be employed to obtain the mapping represented by the DeepONet. The computational cost for obtaining each mapping is comparable as well.

\subsection{Results with arbitrary top and bottom boundaries}

Next, we use the domain mapping method to obtain the mapping in which the bottom boundary is  arbitrary. The bottom boundary is sampled the same as the top one as alluded to earlier, except that some coefficients/parameters are different. Specifically, $\phi_2(a)$ is sampled uniformly from $[0.14,0.24]$, $\phi_2(b)$ uniformly from $[\phi(a)-0.14,\phi(a)+0.14]$, $\lambda \in [0,0.1]$.
Compared to the case where the bottom boundary is flat, we are interested in two issues here: 1. how does the depth between the top and the bottom boundaries affects the flow pattern in the basin? 2. how does the morphology of the bottom boundary affect the flow field in the basin in addition to that of the top boundary?

\begin{figure}[H]
\centering
 \subfigure[$\bh_t=(0.14, 0.14, 0, 0, 0, 0, 0, 0, 0, 0).$]{
			\includegraphics[width=1\textwidth]{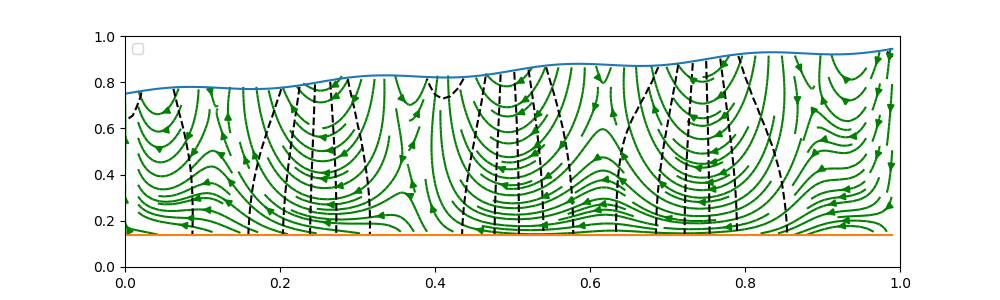}
}\\
 \subfigure[$\bh_t=(0.24, 0.24, 0, 0, 0, 0, 0, 0, 0, 0).$]{
			\includegraphics[width=1\textwidth]{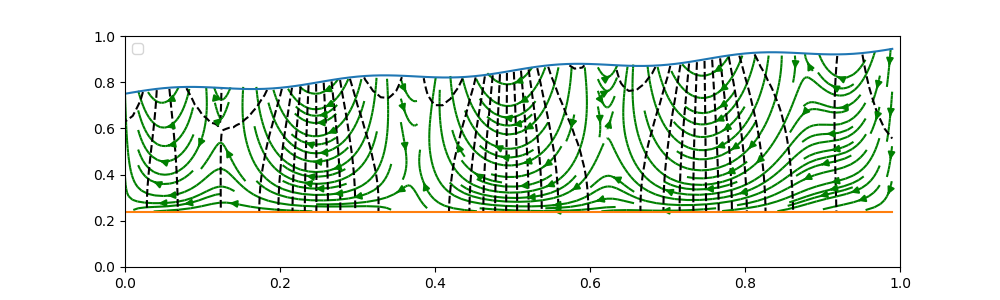}
}
\caption{The steady state solution in the Toth basin with given top and bottom boundaries represented by spectral representations obtained using the domain mapping approach. (a). The flow pattern is altered compared to Figure {\ref{12}}. When the bottom is lifted, i.e., the basin is shallower, the flows are compressed. (b).  When the bottom is lifted further, the flows are compressed further and localization is more prominent. However, the flow pattern does not seem to differ much from (a) qualitatively.   }
\label{110}
\end{figure}

\begin{figure}[H]
\centering
 \subfigure[$\bh_t=(0.14, 0.28, 0, 0, 0, 0, 0, 0, 0, 0).$]{
			\includegraphics[width=0.8\textwidth]{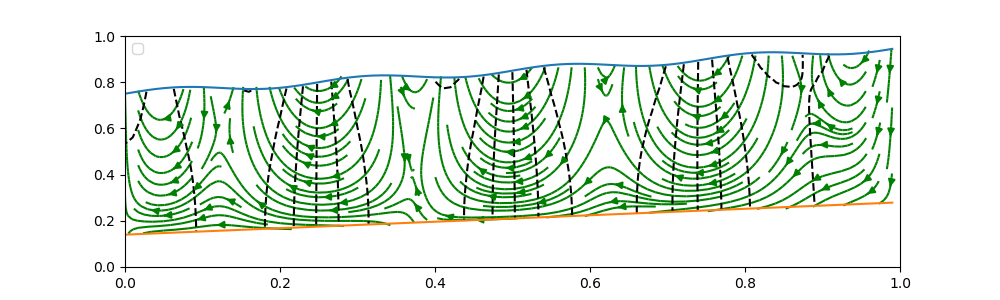}
}\\
 \subfigure[$\bh_t=(0.24, 0.38, 0, 0, 0, 0, 0, 0, 0, 0).$]{
			\includegraphics[width=0.8\textwidth]{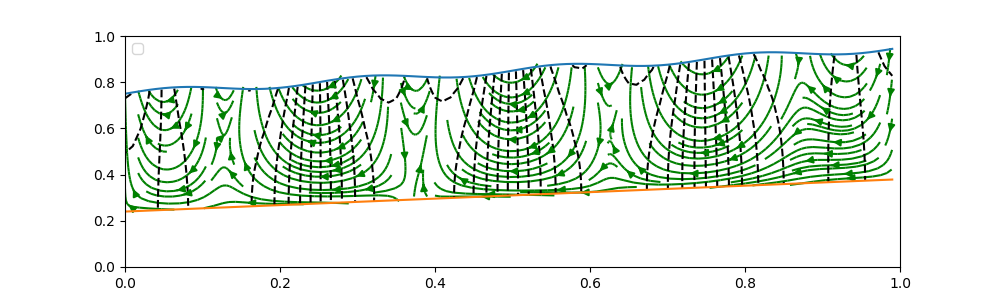}
}\\
 \subfigure[$\bh_t=(0.38, 0.24, 0, 0, 0, 0, 0, 0, 0, 0).$]{
			\includegraphics[width=0.8\textwidth]{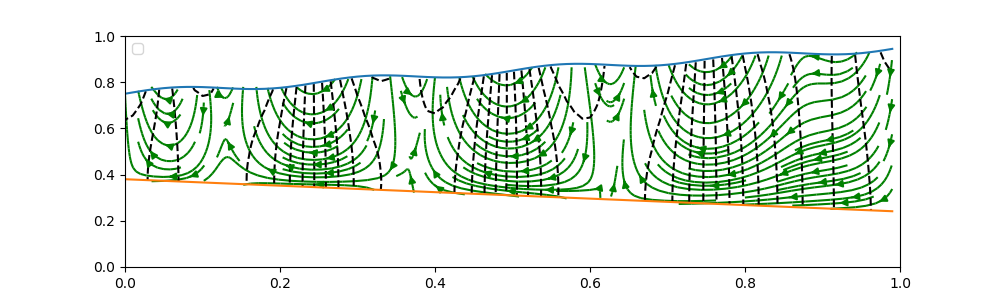}
}\\
 \subfigure[$\bh_t=(0.28, 0.14, 0, 0, 0, 0, 0, 0, 0, 0).$]{
			\includegraphics[width=0.8\textwidth]{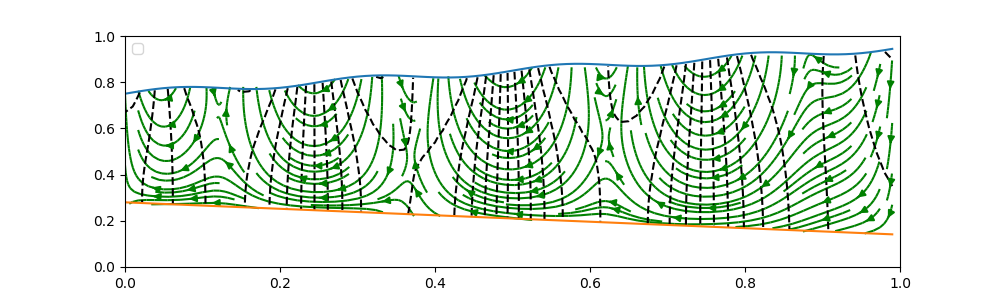}
}
\caption{The steady state solution in the Toth basin with given top and bottom boundaries represented by spectral representations obtained using the domain mapping approach. (a).  The flow has a tendency to flow to the left in long distance due to the increasing slope over there.  (b).  The flow tends to travel long distance down the bottom boundary to the left analogous to (a).  There is no qualitative difference between (a) and (b) where the depth of the basin is different.  (c). Flows are more compartmentalized due to the shallow basin. (d). As the depth increases in the basin, the longer range flow is observed near the bottom.  }
\label{111}
\end{figure}

\begin{figure}[H]
\centering
 \subfigure[$\bh_t=(0.14, 0.28, 0, 0, 0, 0, 0, 0, 0, 0.015).$]{
			\includegraphics[width=0.8\textwidth]{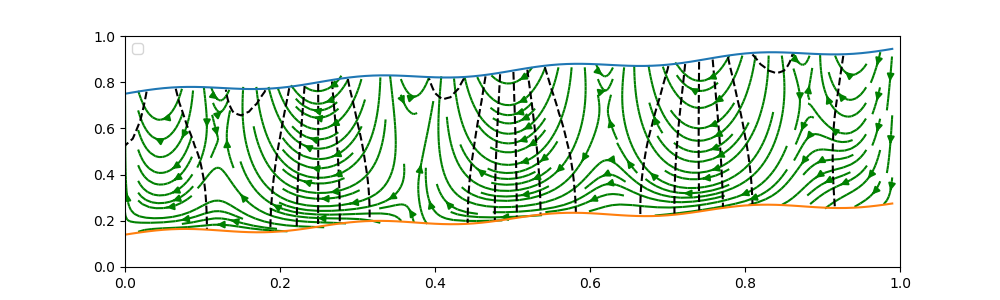}
}\\
\subfigure[$\bh_t=(0.24, 0.38, 0, 0, 0, 0, 0, 0, 0, 0.015).$]{
			\includegraphics[width=0.8\textwidth]{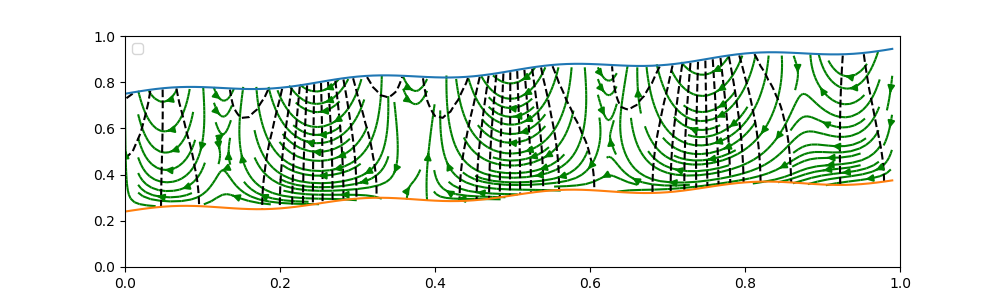}
}
\caption{The steady solution in the Toth basin with wavy top and bottom boundaries represented by spectral representations obtained using the domain mapping approach. (a).  The streamline of the flow is perturbed by the bottom boundary. (b).  A slight  decrease in depth does not make much qualitative difference in the flow pattern.
}
\label{112}
\end{figure}

When the bottom is flat, a decrease in the depth of the basin does not seem to impact much to the overall flow pattern except that the localized/compartmentalized circulation is enhanced at the top and the depth of the circulation region becomes larger in the dimensionless domain as shown in Figure \ref{110}. When the flat bottom boundary is inclined in the same direction as the top boundary does, the local circulatory flow seems to increase near the top boundary as shown in Figure \ref{111}. When the bottom is inclined opposite to that of the top boundary, the increased depth in the far right end alleviates the small scale circulatory motion to a slightly long distance flow pattern across a scale much larger than the previously confined circulatory region (see Figure \ref{111}). When the bottom boundary is wavy, it does not seem to add any new features to the already known flow patterns alluded to earlier. Figure \ref{112} depicts two examples where the bottom boundaries are wavy with different  amplitudes of spatial variations.

\subsection{Robin boundary-value problem}

For the Robin boundary condition given in \eqref{BC3}, we define the loss function  as follows:
\ben
\bea{l}
L(\theta, \bh_t, \epsilon)= \frac{1}{n_{i}}\sum_{j=1}^{n_{i}}(\nabla \cdot K\cdot \nabla G(\bh_t,\epsilon,\bx_j))^2

+\frac{1}{n_{t}}\sum_{j=1}^{n_{t}}[-\gamma(\phi(\bx_j^t)-G(\bh_t,\epsilon, \bx_j^t))+\\\\
\bn\cdot K\cdot \nabla G(\bh_t,\epsilon, \bx_j^t) ]^2
+\frac{1}{n_{l}}\sum_{i=1}^{n_{l}}(G_x(\bh_t,\epsilon,\bx_i^l))^2
+
 \frac{1}{n_{r}}\sum_{i=1}^{n_{r}}( G_x(\bh_t, \epsilon,\bx_i^r))^2
 +\\\\
 \frac{1}{n_{b}}\sum_{i=1}^{n_{b}}( G_y(\bh_t, \epsilon,\bx_i^b))^2
+ [\frac{\Delta x}{3}\big(F(\bx_1, \epsilon)
+4\sum_{i~even \in\{ 2,\cdots,n-1\}}F(\bx_i^t, \epsilon) +\\\\
2\sum_{i~odd \in\{ 3,\cdots,n-2\}} F(\bx_i^t, \epsilon) +F(\bx_n^t, \epsilon) \big)]^2,
\eea
\een
where $\bx_j$ are the interior points and $\bx_j^k$ are boundary points at the top, left, right, and bottom boundary, respectively, $\bn=\frac{1}{\sqrt{1+\phi_x^2}}(-\phi_x, 1)$, and $F(\bx, \epsilon)=G(\bh_t,\epsilon, \bx)-\phi(\bx)$.


The model parameters are the same as those used above. $\gamma$ can be identified as the reciprocal of the  penetration length: the larger $\gamma$ is, the smaller its impact to the flow in the bottom of the basin. Hence, it is expected that the flow pattern should be similar to the case of the Dirichlet boundary condition when $\gamma$ is large.  $\gamma$ is treated as a hyperparameter, which we must set before training to construct the loss function. It could be treated as an input variable for the DeepONet though. But, it would take much longer time to train the neural network in this case. So, we decide not to pursue it in this study. For the Robin boundary condition, we are interested in what changes it brings to the steady flow patterns in comparison to the Dirichlet case.

\begin{figure}[H]
\centering
 \subfigure[$\bh_t= (0.75, 0.95, 0, 0, 0, 0, 0, 0, 0, 0.015).$]{
			\includegraphics[width=1\textwidth]{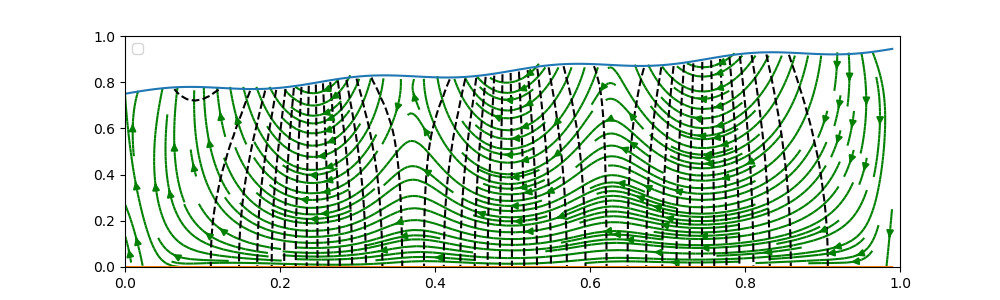}
}
\subfigure[$ \bh_t=(0.75, 0.85, 0, 0, 0, 0, 0, 0, 0, 0.015)$.]{
			\includegraphics[width=1\textwidth]{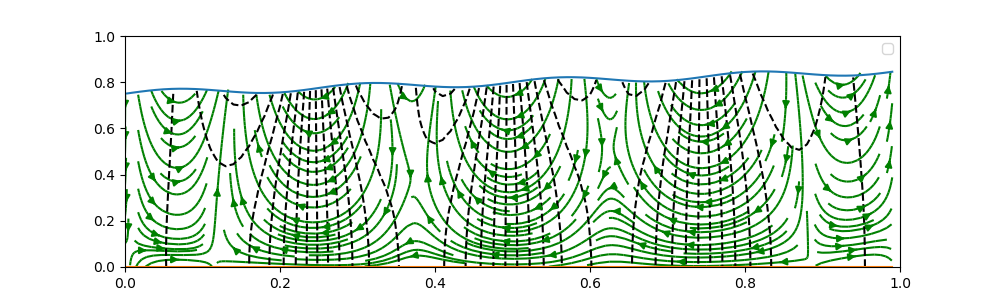}
}
\subfigure[$\bh_t=(0.75, 0.75, 0, 0, 0, 0, 0, 0, 0, 0.015).$]{
			\includegraphics[width=1\textwidth]{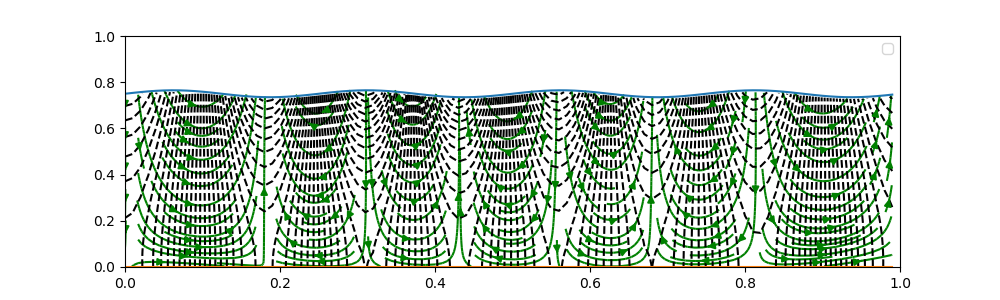}
}
\caption{Steady flow patterns at three selected slopes of the top boundary with $\gamma=10$. (a). The top boundary of a higher slope produces long-distance flow near the impervious bottom boundary. (b). Compartmentalization becomes more prominent as the slope reduces and in the meantime the number of circulatory flow cell increases.   (c) When the average slope of the top surface is zero, long distance flows completely ceases and compartmentalized flow patterns dominate.}
\label{R1}
\end{figure}

\begin{figure}[H]
\centering
 \subfigure[$\bh_t= (0.75, 0.95, 0, 0, 0, 0, 0, 0, 0, 0.015).$]{
			\includegraphics[width=1\textwidth]{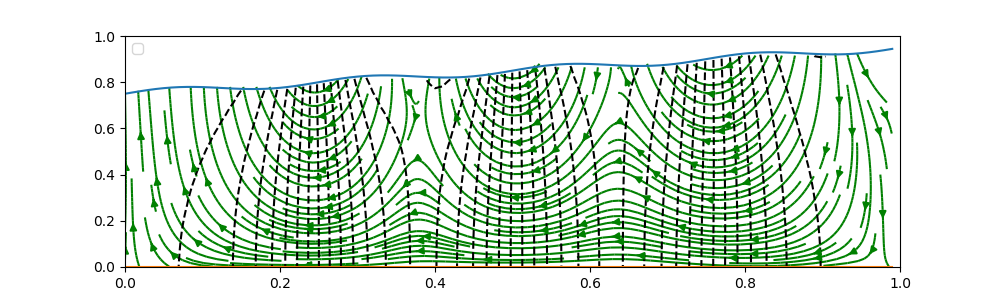}
}
 \subfigure[$\bh_t= (0.75, 0.85, 0, 0, 0, 0, 0, 0, 0, 0.015).$]{
			\includegraphics[width=1\textwidth]{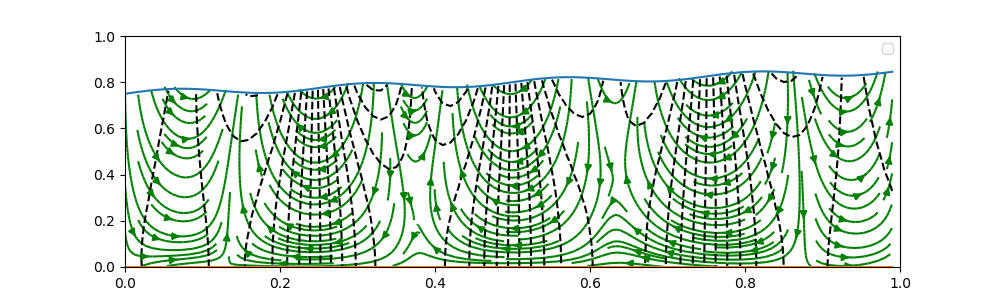}
}
 \subfigure[$\bh_t= (0.75, 0.75, 0, 0, 0, 0, 0, 0, 0, 0.015).$]{
			\includegraphics[width=1\textwidth]{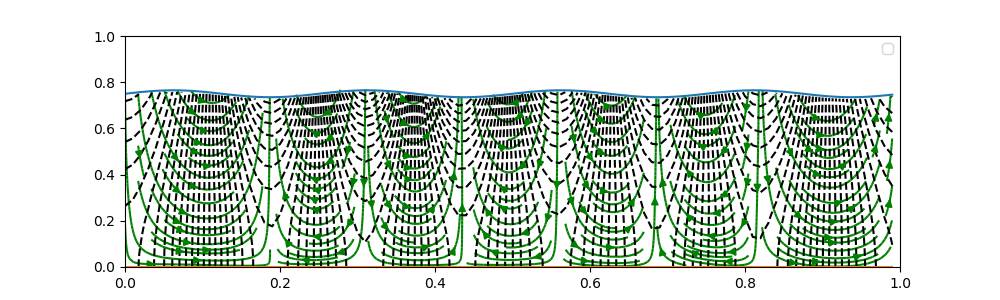}
}
\caption{Steady flow patterns at three selected slopes of the top boundary with $\gamma=10^7$. (a). The top boundary of a higher slope produces long-distance flows near the impervious bottom boundary. (b). Compartmentalization becomes more prominent as the slope reduces and in the meantime the number of circulatory flow cell increases.   (c) When the average slope of the top surface is zero, the long distance flow completely ceases and compartmentalized flow patterns dominate. The results are similar to those in Figure \ref{R1}.}
\label{R2}
\end{figure}

By examining the numerical results, we observe that when $\gamma\ge 10$, the flow patterns in the three cases with different slopes and spatial variations are pretty much independent of the increase in values of $\gamma$. Namely, the patterns in Figure \ref{R1} ($\gamma=10$) are nearly the same  as those in Figure \ref{R2} when $\gamma=10^{7}$. While $\gamma$ becomes smaller however, the flow patterns change quite dramatically, the number of localized flow patterns decreases and long distance flows become more prominent. This is because the penetration length is enhanced in this case so that flow patterns near the bottom are affected directly by the top boundary condition.  We expect that the solution is going to be approaching the solution of the Dirichlet problem as $\gamma \to \infty$. The steady states we have calculated do support the observation.

\begin{figure}[H]
\centering
 \subfigure[$\bh_t=(0.75, 0.95, 0, 0, 0, 0, 0, 0, 0, 0.015).$]{
			\includegraphics[width=1\textwidth]{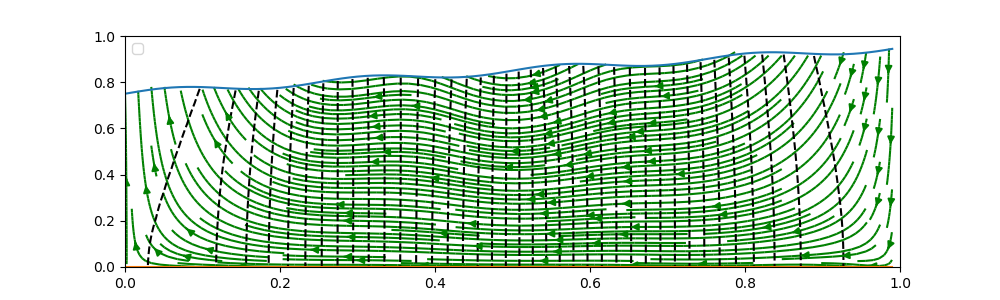}
}
\subfigure[$\bh_t= (0.75, 0.85, 0, 0, 0, 0, 0, 0, 0, 0.015).$]{
			\includegraphics[width=1\textwidth]{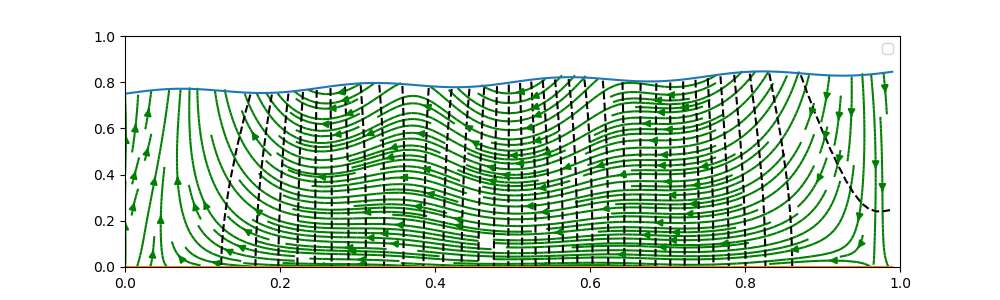}
}
\subfigure[$\bh_t= (0.75, 0.75, 0, 0, 0, 0, 0, 0, 0, 0.015).$]{
			\includegraphics[width=1\textwidth]{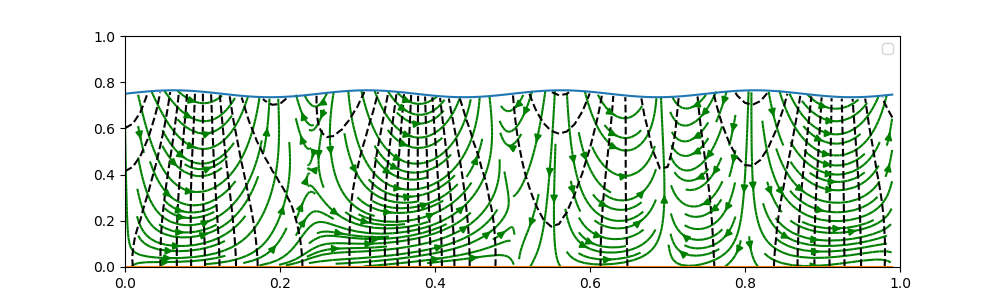}
}
\caption{Steady flow patterns at three selected slopes of the top boundary with $\gamma=0.2$. (a). The top boundary of a higher slope produces long-distance flows near the impervious bottom boundary. (b). Compartmentalization appears near the top locally.   (c) When the average slope of the top surface is zero, the long distance flow across the entire bottom of the basin ceases. However, the compartmentalized flow cells are larger than in the case where $\gamma\geq 10.$ }
\label{}
\end{figure}

\section{Remark on  inhomogeneous equations and time dependent groundwater flow equations}

The method extends readily to the boundary value problem with a source/sink term $Q$ and the initial boundary value problem of \eqref{eq1} in the framework of PIML. For the inhomogeneous boundary value problem, we modify the loss function by including source term $Q$ is the penalty terms in the interior equation and the consistency condition, respectively. For the time-dependent initial-boundary value problem, we add one more dimension in time and sample randomly in both space and time for the interior and boundary points. The changes that need to be made are in the interior penalty term and the consistency condition term in the loss function. The $L_2$ norms used in the loss function needs o be expanded to include the average in time. Some additional techniques in training such as time-domain decomposition, etc. need to be considered to accelerate the training process \cite{Lu2021LearningNO}.

 \section{Conclusion}
We have introduced a new approach to obtaining a mapping between surface topography and the solution of the steady state groundwater flow equation in Toth basins with arbitrary surface topographies. This method utilizes PIML with DeepONet and relies on inputting parameters, such as the conductivity tensor and aspect ratio, into the groundwater flow equation. The resulting boundary-to-solution mapping can be learned and repeatedly used to map out the underground steady state flow field for any surface topography and impervious boundary of any shape, providing an alternative means to study groundwater flow phenomena in complex geophysical systems. This method can estimate groundwater flow patterns in new Toth basins with similar geophysical parameters, and new models can be efficiently machine-learned through transfer learning with a predetermined architecture even in locations where geophysical parameters differ. With additional computational efforts, the model parameters can also be treated as an input to the DeepONet, expanding the mapping to a wider range of groundwater flow equations with varying parameter values. As a result, the DeepONet can be used as a well-trained "knowledgeable" boundary-to-solution predictor, demonstrating a neural network representation of an inverse differential operator in hydrological applications, which can be applied in numerous cases.

\subsection*{Acknowledgements}

Jun Li's work is partially supported by The Science \& Technology Development Fund of Tianjin Education Commission for Higher Education 2020KJ005, and the National Natural Science Foundation of China 11971247. Qi Wang's work is partially supported by NSF DMS-1954532.


\bibliographystyle{unsrt}

\end{document}